\documentstyle[12pt]{article}
\catcode`@=11
\newcount\@tempcntc
\def\@citex[#1]#2{\if@filesw\immediate\write\@auxout{\string\citation{#2}}\fi
  \@tempcnta\z@\@tempcntb\m@ne\def\@citea{}\@cite{\@for\@citeb:=#2\do
    {\@ifundefined
       {b@\@citeb}{\@citeo\@tempcntb\m@ne\@citea\def\@citea{,}{\bf ?}\@warning
       {Citation `\@citeb' on page \thepage \space undefined}}%
    {\setbox\z@\hbox{\global\@tempcntc0\csname b@\@citeb\endcsname\relax}%
     \ifnum\@tempcntc=\z@ \@citeo\@tempcntb\m@ne
       \@citea\def\@citea{,}\hbox{\csname b@\@citeb\endcsname}%
     \else
      \advance\@tempcntb\@ne
      \ifnum\@tempcntb=\@tempcntc
      \else\advance\@tempcntb\m@ne\@citeo
      \@tempcnta\@tempcntc\@tempcntb\@tempcntc\fi\fi}}\@citeo}{#1}}
\def\@citeo{\ifnum\@tempcnta>\@tempcntb\else\@citea\def\@citea{,}%
  \ifnum\@tempcnta=\@tempcntb\the\@tempcnta\else
   {\advance\@tempcnta\@ne\ifnum\@tempcnta=\@tempcntb \else \def\@citea{--}\fi
    \advance\@tempcnta\m@ne\the\@tempcnta\@citea\the\@tempcntb}\fi\fi}
\catcode`@=12

\def\theequation{\arabic{section}.\arabic{equation}}
\begin{document}

\begin{flushright}
RAL-TR/95-035\\
August 1995
\end{flushright}

\begin{center}
{\bf{\Large {\em W}- and {\em Z}-Boson Interactions in
Supersymmetric}}\\[0.4cm]
{\bf{\Large Models with Explicit {\em R}-Parity Violation }}\\[2.4cm]
{\large Marek Nowakowski}$^a$\footnotemark[1]
{\large and Apostolos Pilaftsis}$^b$\footnote[1]{E-mail addresses:
nowakowski@lnf.infn.it and pilaftsis@v2.rl.ac.uk}\\[0.4cm]
$^a${\em INFN-Laboratori Nazionali di Frascati, P.O.\ Box 13, I-00044
Frascati (Roma), Italy}\\[0.3cm]
$^b${\em Rutherford Appleton Laboratory, Chilton, Didcot, Oxon, OX11 0QX, UK}
\end{center}
\vskip1.5cm
\centerline{\bf ABSTRACT}
The minimal supersymmetric Standard Model with explicit $R$-parity
nonconservation contains bilinear terms involving the left-handed lepton
superfields and the Higgs chiral multiplet with hypercharge $Y=+1$, which
cannot in general be rotated away in the presence of
soft-supersymmetry-breaking interactions. These bilinear
lepton-number-violating terms are found to give rise to non-zero vacuum
expectation values of the scalar neutrinos. This leads to nonuniversal and
flavour-violating tree-level couplings of the $W$ and $Z$ bosons to charged
leptons and neutrinos. The parameter space of this novel scenario is
systematically analyzed and further restricted by a number of laboratory,
astrophysical, and cosmological constraints. The possibility that our minimal
model can account for the KARMEN anomaly is examined.

\newpage

\section{Introduction}
\indent

The minimal supersymmetric Standard Model (MSSM) with the particle
content of the Standard Model (SM), including their supersymmetric (SUSY)
partners, conserves the discrete quantum number $R$, known as $R$ parity.
Under this symmetry, the SM particles are even whereas their superpartners
are odd. The quantum number of $R$ parity may conveniently be expressed
as~\cite{Rsusy,SW,HS,EGJRV}.
\begin{equation} \label{m1}
R=(-1)^{3B+L+2S} \, ,
\end{equation}
where $B$, $L$, and $S$ are the baryon number, the lepton number, and the
spin of a particle, respectively. Evidently, nonconservation of $R$ parity
results, in general, in baryon ($\not\!\! B$) and lepton ($\not\!\! L$)
number violating terms. However, the usually considered models retain only
the lepton-number-violating terms in order to preserve the stability of the
proton. In the presence of both, $\not\!\! L$ and $\not\!\! B$ terms, the
proton decay might render the model phenomenologically not
viable.

The explicit breaking of $R$ parity in a SUSY model means also that we
introduce additional coupling constants into the Lagrangian, but not additional
fields, {\em i.e.}\ the field content of the model is the same as in the MSSM.
This property of the model has some interesting consequences. One important
implication is that the explicit lepton-number violation allows, in
principle, for a mixing of the left-handed neutrinos with the gauginos
and higgsinos. As a consequence, such a mixing, if present,
will generate neutrino masses. This mechanism to render neutrinos massive
is quite distinctive from the standard options, such as the introduction of
right-handed neutrinos or the extension of the Higgs sector by adding an
exotic Higgs triplet. However, to make the afore-mentioned mixing possible,
one should ensure that the sneutrinos acquire vacuum expectation values
(vev's).
By a careful examination of the scalar potential of the $R$-parity broken
SUSY model, one can show that the appearance of non-vanishing vev's
for the scalar neutrinos  is not always a direct consequence of the
theory. In fact, if a term of the form
$\varepsilon_i \varepsilon_{ab}\hat{L}^a_i \hat{H}^b_1$ is neglected
in the superpotential of the model, where $\hat{L}_i$ are the chiral lepton
fields and $\hat{H}_1$ is the super-Higgs fields with hypercharge $Y=1$, then
insistence of non-zero vev's of the sneutrino fields can lead, in some cases,
to a fine-tuning relation among the original parameters of the scalar
potential. For instance, this would happen if we assumed the scalar mass
terms $m_{ij}L^{\dagger}_iL_j +$ H.c.\ to be diagonal in $i$, $j$.
On the other hand,
the retention of the $\varepsilon_i$ terms leads unavoidably to non-zero vev's
of the sneutrinos ($w_i$). It is also easy to see that in the presence of
soft-SUSY breaking parameters, the $\varepsilon_i$ terms cannot be rotated
away by a unitary transformation. This is consistent with the observation
that such a term will be generated radiatively, even if one chooses
$\varepsilon_i=0$ at the Planck mass scale~\cite{HS}.
Therefore, one could argue that models which break the $R$ parity explicitly,
in which the $\varepsilon_i$ terms are neglected, may be considered to be not
general and hence incomplete. This fact is expected to influence the analytic
expressions of the mass and mixing matrices, especially when new parameters
enter the theory.  Indeed, we will show how the $\varepsilon_i$ terms
change the phenomenological predictions of the model under consideration.

Models with an explicit mixing term between $\hat{L}_i$ and $\hat{H}_1$ have
been considered in the past~\cite{HS}. More recently, such models have been
re-examined by paying special attention to CP violation in the scalar potential
as well as to neutrino masses \cite{JN,JNCP}.
In Refs.~\cite{HS,JN,JNCP}, it was assumed that all
individual lepton numbers, $L_i$, are broken. It has been known for some time
that lepton-number-violating interaction can erase the existing baryon
asymmetry in the universe (BAU) if $\not\!\! L$ interactions are
in thermal equilibrium with the $B+L$-violating sphalerons.
The most stringent constraints on $w_i$ and $\varepsilon_i$ arise from such
considerations. However, it has been realized~\cite{DR}
that one can evade the erasure
of the BAU if we demand that one individual lepton number is conserved.
No constraints on lepton-number-violating couplings will then follow. The
most conservative constraints on lepton-number-violating couplings can then
be inferred from an analysis of the absence of possible
new-physics phenomena in various experiments.

In $R$-parity broken models, there are in general two sources that can
produce lepton-number- and lepton-flavour-violating interactions. Those that
are
induced by the $\varepsilon_i$ and $w_i$ parameters and give rise, {\em
 e.g.}, to tree level off-diagonal $Z$-boson decays, and those that are
proportional to the so-called $\lambda$ and $\lambda'$ couplings in
the superpotential. The literature pertinent to constraints on the
$\lambda$ and $\lambda'$ couplings is immanent~\cite{hinch,BGH,GBDC,HKK,roh}.
Here, we will study effects of lepton-number and/or lepton-flavour
violation on the ordinary known matter, which can be induced by $W$-
and/or $Z$-boson interactions in an $R$-parity broken model. Therefore,
one may consider our study complementary to investigations of limits
on the $\lambda$ and $\lambda'$ couplings mentioned above. In an
$\not\!\! R$ SUSY model, direct limits on $w_i$ are mostly obtained
from upper bounds on light neutrino masses, {\em e.g.}\
using the laboratory bound $m_{\nu_{\tau}} < 31$~MeV~\cite{PDG94}.
However, we find that better limits may be deduced
by the nonobservation of non-SM processes, such as the decay of a muon
into three electrons. Moreover, cosmological and astrophysical
implications of a massive light neutrino for our model will be
discussed in Section~5.7.

The paper is organized as follows. In Section 2, we will give a detailed
discussion of the $R$-parity broken SUSY model where the emphasis
will be put on the scalar potential of the model.
Further analytic results on this topic are relegated in Appendix A.
Section 3 treats the mixing between the neutralinos and
neutrinos as well as that between the charginos and charged leptons.
In Section 4, we derive the
$W$- and $Z$-boson interaction Lagrangians in the seesaw approximation.
In Section 5,
we analyze a number of low-energy processes that can be induced by the
non-SM couplings present in $\not\!\! R$ SUSY models. We then discuss
new constraints that may be derived by laboratory experiments together
with constraints coming from cosmology and astrophysics. In addition,
we investigate the possibility that our $R$-parity broken SUSY model
can explain the KARMEN anomaly. Our numerical results are presented in
Section 6. We draw our conclusions in Section 7.

\setcounter{equation}{0}
\section{The $R$-parity violating SUSY model}
\indent

In this section, we set up our definition and notation and outline
in some more detail the scalar potential of the MSSM with explicit
$R$-parity breaking terms. Our main concern will be to argue that
the retention of a special term in the superpotential, which is usually
not taken into account, leads {\it naturally}, {\em i.e.}\ without
any fine tuning problems, to non-zero vev's of the sneutrinos.
As a consequence, neutrinos acquire masses through mixing with
gauginos and higgsinos and the standard $W$/$Z$-boson interactions with
fermions get modified, {\em e.g.}\ one gets tree-level off-diagonal $Z$
decays.
The novelty in our approach is the afore-mentioned term in the superpotential
whose coupling strength $\varepsilon_i$ enters the $W$/$Z$ interaction
Lagrangians.

\subsection{Superpotential}
\indent

We write the full superpotential $W$ as consisting of an $R$-parity
conserving part ($W_0$) and $R$-parity violating term
($W_{\not \! R}$), {\em i.e.}\
\begin{equation} \label{m2}
W\ =\ W_0+W_{\not \! R}\, .
\end{equation}
Let then $\hat{L}_i$ ($\hat{E}_i^C$) and
$\hat{Q}_i$ ($\hat{U}_i^C $,$ \hat{D}_i^C$) denote the lepton and
quark doublets superfields (lepton and quarks $SU(2)$ singlets) with
generation index $i$, respectively and let $\hat{H}_{1,\; 2}$ be
the super-Higgs fields. With the usual $U(1)_Y$ quantum
number assignment, $Y(\hat{L}_i)=-1$,
$Y(\hat{E}_i^C)=2$, $Y(\hat{Q}_i)=1/3$, $Y(\hat{D}_i^C)=2/3$,
$Y(\hat{U}_i^C)=-4/3$, $Y(\hat{H}_1)=-1$, $Y(\hat{H}_2)=1$,
the standard form for $W_0$ is
\begin{equation} \label{m3}
W_0\ =\ \varepsilon_{ab}\left [h_{ij}\hat{L}_i^a \hat{H}_1^b
\hat{E}_j^C + h'_{ij}\hat{Q}_i^a \hat{H}_1^b \hat{D}_j^C + h''_{ij}
\hat{Q}_i^a \hat{H}_2^b \hat{U}_j^C + \mu \hat{H}_1^a \hat{H}_2^b \right]\, ,
\end{equation}
where $a$, $b$ are $SU(2)$ group indices.

The explicit breaking of $R$ parity can be introduced through
$W_{\not \! R}$, which in its most general form is given by~\cite{SW,HS}
\begin{equation} \label{m4}
W_{\not \! R}\ =\ \varepsilon_{ab}\left( \lambda_{ijk}\hat{L}_i^a
\hat{L}_j^b
\hat{E}_k^C + \lambda'_{ijk}\hat{L}_i^a \hat{Q}_j^b \hat{D}_k^C
+ \varepsilon_i \hat{L}_i^a \hat{H}_2^b \right) + \lambda''_{ijk}
\hat{U}_i^C \hat{D}_j^C \hat{D}_k^C \, .
\end{equation}
Unlike the MSSM, the $R$-parity broken SUSY model allows for explicitly
broken lepton ($\not \! \! L$) and baryon ($\not \! \! B$) number
interactions. To be more precise, the terms in Eq.~(\ref{m3}) proportional to
$\lambda_{ijk}=-\lambda_{jik}$, $\lambda'_{ijk}$ and $\varepsilon_i$  violate
lepton number, whereas the baryon number is explicitly broken by the
$\lambda''_{ijk}$-term ($\lambda''_{ijk}=-\lambda''_{ikj}$).
The presence of both $\not\!\! B$- and $\not\!\! L$-type of terms in
the Lagrangian leads to unsuppressed proton decay.
Therefore, at the most, we can retain either the
$L$-violating or the $B$-violating terms in (\ref{m3}).
To account for this fact, hereafter we will set $\lambda''_{ijk}=0$.

Let us now comment on the term
$\varepsilon_{ab} \varepsilon_i \hat{L}_i^a \hat{H}_2^b$ which is usually
not taken into account in $W_{\not \! R}$ by using the argument of
field redefinitions to rotate away such bilinears. It has been discussed
in some detail in~\cite{JN,JNCP}
that this argument may not be valid once
we add to the Lagrangian soft-SUSY breaking terms.
Indeed, we are unable to absorb the $\varepsilon_i$ terms by using an
orthogonal
transformation of the $\hat{H}_1$ and $\hat{L}_i$ fields. The omission of
such terms is therefore not justified. We note here that exactly these terms
are in principle responsible for a non-zero vev's of the sneutrinos.
The $\varepsilon_i$ terms force the
vev's of the sneutrino fields to assume non-zero values. We will
return to this point while discussing the Higgs potential.

The $\lambda$-terms in $W_{\not \! R}$ lead to the
interaction Lagrangian
\begin{equation} \label{Llambda}
{\cal L}_{\lambda}\ =\ \lambda_{ijk}\biggl[\tilde{\nu}_{iL}
\bar{e}_{jR}e_{jL}+
\tilde{e}_{jL}\bar{e}_{kR}\nu_{iL}+\tilde{e}^*_{kR}
\bar{\nu}_{iL}^Ce_{jL}
-(i \leftrightarrow j)\biggr]\ +\ \mbox{H.c.}\, ,
\end{equation}
where $e_i$ denote charged leptons enumerated by generation index
$i$ and tilded symbols denote as usual the superpartners. Correspondingly,
the $\lambda'$ interaction Lagrangian reads
\begin{eqnarray} \label{Llambda'}
{\cal L}_{\lambda'}&=&\lambda'_{ijk}\biggl( \tilde{\nu}_{iL}
\bar{d}_{kR}d_{iL}
+\tilde{d}_{jL}
\bar{d}_{kR}\nu_{iL}+\tilde{d}^*_{kR}\bar{\nu}_{iL}^Cd_{jL}
- \tilde{e}_{iL}\bar{d}_{kR}u_{jL}
- \tilde{u}_{jL}\bar{d}_{kR}e_{iL}\nonumber\\
&& -\tilde{d}^*_{kR}\bar{e}_{iL}^Cu_{jL}
\biggr)\ +\ \mbox{H.c.}\, ,
\end{eqnarray}
where $d_i$ ($u_i$) are down-type (up-type) quarks.

Since the main concern of the present paper will be to constrain the
lepton-number-violating couplings,
it is worth discussing briefly constraints on
the $L$-violating couplings which come from baryogenesis. It has been known
for some time that non-perturbative anomalous $B$-violating interactions
of the electroweak theory can
wash out the baryon asymmetry generated initially at the GUT scale. This gives
rise to severe constraints on lepton-number-violating couplings, such as
the trilinear couplings $\lambda_{ijk}$ and $\lambda'_{ijk}$.
In this context, it has been noticed that to evade such limits is sufficient
to have one individual lepton number, $L_i$, conserved~\cite{DR}.
The latter will be assumed throughout this work. In particular, we have
\begin{eqnarray} \label{m7}
&&\lambda_{ijk}=0,\quad \mbox{for} \quad i\neq j \neq k\, , \nonumber \\
&&\lambda_{iki} \neq 0,\quad \mbox{for}\quad \not\! \! L_k \,\,\, L_i\, ,
\nonumber \\
&&\lambda'_{ijk} \neq 0,\quad \mbox{for }\quad \not\! \! L_i\, ,
\end{eqnarray}
where~$\not\! \!  L_i$ ($L_i$) indicates which lepton number is violated
(conserved). Choosing $L_i$ as the conserved lepton number, we then have
four independent $\lambda$ couplings. For instance, conserving $L_e$, we
are left with non-zero $\lambda_{121}$, $\lambda_{131}$, $\lambda_{232}$, and
$\lambda_{323}$. This conservation of a separate lepton number,
{\em i.e.}\ $\Delta L_e=0$,
has profound consequences for exotic lepton-number-violating processes, which
can, in principle, occur at the tree level in an $R$-parity broken SUSY model.
If one $L_i$ is strictly conserved, then no tree-level $\lambda$-dependent
interaction can contribute, {\em e.g.}, to the process $\mu \to eee$.
The latter would proceed via a sneutrino mediated diagram only if {\em all}
individual lepton numbers were broken. If $\Delta L_i=0$ for only one
lepton number $L_i$, the bound on the $\lambda$ couplings derived
via $\mu \to eee$ in \cite{hinch} does not further apply. The reason is
that the process $\mu \to eee$ is still possible at the tree level,
however through a diagram containing off-diagonal $Z$ couplings
(see discussion in Section~5.1).
In the subsequent section, we will also address the question whether
having $\Delta L_i=0$ for one lepton number $L_i$ at the level of Lagrangian,
this particular symmetry gets broken spontaneously through a vev of the
sneutrino field, $w_i$, with the very same flavour index $i$. Indeed, we
will argue that this is not the case, when the couplings $\varepsilon_i$
are taken into account.

Laboratory constraints on $\lambda$ and $\lambda'$ couplings have been put
in~\cite{BGH} and~\cite{GBDC,HKK}. The detectability of possible direct
$R$-parity-violating signals through the $\lambda$ and $\lambda'$
couplings at the CERN Large Electron Positron (LEP) collider, planned
to operate at 200-GeV centre of mass energies (LEP-2), has been discussed
in~\cite{roh}.

We note here that there are low-energy processes to which both type of
$\not\!\! L$ interactions ---those induced by the trilinear $\not\!\! R$
$\lambda$- and $\lambda'$-dependent couplings and those emanating from
non-zero $\varepsilon_i$ and $w_i$ parameters--- will contribute. We will
see that a combined analysis is not necessary, {\em i.e.}\
deriving limits first on $\varepsilon_i$ and $w_i$ from, say,
$\mu \to eee$, and then applying the so obtained results to put constraints
on the $\lambda$ and $\lambda'$ couplings.

\subsection{The scalar potential}
\indent

With the superpotential given, it is straightforward to construct the scalar
potential of an $R$-parity broken SUSY model~\cite{Rsusy}.
Using the convention that
symbols without a hat, say $A$, are the spin-zero content of a chiral
superfield
$\hat{A}$, we first write down the relevant soft-SUSY breaking terms
\begin{eqnarray} \label{m8}
V_{soft}&=&m_1^2 H_1^{\dagger}H_1 + m_2^2 H_2^{\dagger} H_2
+(m_{L_{ij}}^2L_i^{\dagger}L_j\ +\  \mbox{H.c.})\nonumber \\
&&-(m_{12}^2\varepsilon_{ab}H_1^a H_2^b\ +\ \mbox{H.c.})
+ (\kappa'_i \varepsilon_{ab}
H_2^a L_i^b\ +\ \mbox{H.c.}) \nonumber \\
&&+ (\mu^2_{+ij}E_i^* E_j\ +\ \mbox{H.c.})+[\mu'_{ij}(H_1^{\dagger}L_i)E_j\ +\
\mbox{H.c.}] \nonumber \\
&&+ (\kappa'_{ijk}\varepsilon_{ab}L^a_j L^b_j E_k\ +\ \mbox{H.c.})\, .
\end{eqnarray}
In Eq.~(\ref{m8}), $E_i$ are the positively charged scalar singlet fields.

The full scalar potential (without squarks) can be written as the sum of five
terms,
\begin{equation} \label{m9}
V_{Scalar}\ =\ V^{2H} + V^L + V^{\not L} + V_+^L + V_+^{\not L}\, ,
\end{equation}
where $V^{2H}$ is the usual Higgs potential of the MSSM, $V^L$ and $V
^{\not L}$
contain the slepton doublets (both $L$-conserving and $L$-violating parts, the
latter connected with $R$-parity breaking couplings), and finally
$V_+^L$  and $V_+^{\not L}$
refers to the part of the potential which contains
the charged scalars $E_i$. The standard MSSM potential, $V^{2H}$, reads
\begin{eqnarray} \label{m10}
V^{2H}&=&\mu_1^2\phi_1^{\dagger}\phi_1 +
\mu_2^2 \phi_2^{\dagger} \phi_2
+{1 \over 2}\lambda_1(\phi_1^{\dagger}\phi_1)^2 +
{1 \over 2}\lambda_1(\phi_1^{\dagger}
\phi_2)^2
\nonumber \\
&&+\lambda_3 (\phi_1^{\dagger}\phi_1)(\phi_2^{\dagger}\phi_2)
-(\lambda_3 + \lambda_1)(\phi_1^{\dagger}\phi_2)(\phi_2^{\dagger}\phi_1)
\nonumber \\
&&+ \lambda_6 (\phi_1^{\dagger}\phi_2)
+ \lambda_6^*(\phi_2^{\dagger}\phi_1)\, ,
\end{eqnarray}
where we have used the notation
$\varphi_i \equiv L_i$, $\phi_2
\equiv H_2$, and $\phi_1 \equiv -i\tau_2 H_1^*$ [$\tau_2$ being
the Pauli matrix, ($i\tau_2)_{ab}=\varepsilon_{ab}$]. The new parameters in
Eq.~(\ref{m10}), which depend on the couplings entering the superpotential,
are the soft-SUSY breaking parameters given in (\ref{m8}), as well as the
$SU(2)_L$  coupling constant $g$ and the corresponding $U(1)_Y$ coupling
constant $g'$. These parameters are related as follows:
\begin{eqnarray} \label{m11}
&&\mu_1^2=m_1^2+\vert \mu \vert^2,\; \; \; \mu_2^2=m_2^2 +\vert
\mu \vert^2 + \varepsilon_i \varepsilon_i^*\, , \nonumber \\
&&\lambda_1={1 \over 4}(g^2 + g'^2), \; \; \; \lambda_3={1 \over
2}g^2- \lambda_1,\; \; \;  \lambda_6=-m_{12}^2\, .
\end{eqnarray}
The lepton-number-conserving scalar potential containing the slepton doublet
fields $\varphi_i$ reads
\begin{eqnarray} \label{m12}
V^L &=& (\mu^2_{L_{ij}}\varphi_i^{\dagger}\varphi_j\ +\ \mbox{H.c.})
+ {1 \over 2}\lambda_1 (\sum_i \varphi_i^{\dagger}\varphi_i)^2 +
\lambda_1(\phi_1^{\dagger}\phi_1)(\varphi_i^{\dagger}\varphi_i)
\nonumber \\
&&- \lambda_1(\phi_2^{\dagger}\phi_2)(\varphi_i^{\dagger}\varphi_i)
+ (\lambda_3 + \lambda_1)(\phi_2^{\dagger}\varphi_i)(\varphi_i^{\dagger}
\phi_2) \nonumber \\
&&+ \Big[\kappa_{jk} -(\lambda_3 + \lambda_1)\delta_{jk}\Big](\phi_1^{\dagger}
\varphi_k)(\varphi_j^{\dagger}\phi_1) + \kappa_{nmij}(\varphi_i^T \tau_2
\varphi_j)(\varphi_n^T \tau_2 \varphi_m)^{\dagger}\, .
\end{eqnarray}
Here, we have used the definitions
\begin{eqnarray} \label{m13}
&&\mu_{L_{ij}}^2=m^2_{L_{ij}}+\varepsilon^*_i \varepsilon_j, \,\,\,
\kappa_{jk}=\kappa^*_{kj}\equiv h^*_{ji}h_{ki}, \nonumber \\
&&\kappa_{nmij}=-\kappa_{mnij}=-\kappa_{nmji}=\kappa^*_{ijnm} \equiv
\lambda^*_{nmk}\lambda_{ijk}\, .
\end{eqnarray}
$V^L$ given in Eq.~(\ref{m12}) may be contrasted with the corresponding
potential of the MSSM. In the MSSM, the term
proportional $\kappa_{nmij}$ is absent and
$\mu^2_{L_{ij}}$ should be replaced by the diagonal
mass parameters $\mu^2_{L_i}=m^2_{L_i}+\vert \varepsilon_i \vert^2$.

Finally, the lepton-number-violating part of the potential is given by
\begin{eqnarray} \label{m14}
V^{\not L}&=&[i\kappa_i (\phi_1^T \tau_2 \varphi_i)\ +\ \mbox{H.c.}]
+ [i\kappa'_i (\phi_2^T \tau_2 \varphi_i)\ +\ \mbox{H.c.}] \nonumber \\
&&- i\kappa_{nmj}(\phi_1^{\dagger}\varphi_j)(\varphi_n^T \tau_2
\varphi_m)^{\dagger}\, ,
\end{eqnarray}
with
\begin{equation} \label{m15}
\kappa_i \equiv \mu^* \varepsilon_i, \,\,\, \kappa_{nmj} \equiv
\lambda^*_{nmk}h_{jk}\, ,
\end{equation}
and $\kappa'_j$ is a soft-SUSY breaking parameter from Eq.~(\ref{m8}). The
parts
$V^{2H}$, $V^L$ and $V^{\not L}$
are sufficient to determine the minimization conditions of
the potential. The explicit form of the remaining
contributions to $V_{Scalar}$, $V_+^L$ and $V_+^{\not  L}$, is given in
Appendix A. At the minimum, the fields take the values
\begin{equation} \label{m16}
\langle \phi_1 \rangle =\left(\begin{array}{c}
0 \\
v_1 \end{array}\right), \,\,\,
\langle \phi_2 \rangle =\left(\begin{array}{c}
0 \\
v_2 \end{array}\right), \,\,\,
\langle \varphi_i \rangle =\left(\begin{array}{c}
w_i \\
0 \end{array}\right),
\end{equation}
and the minimization conditions are found to be
\begin{eqnarray} \label{m17}
\mu_1^2 v^*_1 +\lambda_1 v_1^*(\vert v_1 \vert^2 -\vert v_2 \vert^2 +
\sum_k \vert w_k \vert^2)+\lambda_6^* v^*_2 -\sum_k \kappa_k w_k &=& 0\, ,
\nonumber \\
\mu_2^2 v^*_2 -\lambda_1 v_2^*(\vert v_1 \vert^2 -\vert v_2 \vert^2 +
\sum_k \vert w_k \vert^2)+\lambda_6 v^*_1 -\sum_k \kappa'_k w_k &=& 0\, ,
\nonumber \\
\sum_j \mu^2_{L_{ij}}w^*_j +\lambda_1 w_i^*(\vert v_1 \vert^2 -\vert v_2
\vert^2 +\sum_k \vert w_k \vert^2)- \kappa_i v_1-\kappa'_i v_2 &=& 0\, ,
\end{eqnarray}
where the last equation is valid for every generation index $i$. It is clear
from Eq.~(\ref{m17}) that $w_i \neq 0$ unless $\kappa_i=\kappa'_i=0$.

To further elucidate this point,
let us first consider the case of $\kappa_i$ and
$\kappa'_i$ not identical to zero ({\em i.e.}\ the case of
the $R$-parity broken SUSY model with $\varepsilon_i \neq 0$),
and assume $w_i =0$ for some generation index $i$.
{}From Eq.~(\ref{m17}), we then get a `fine-tuning' relation among the
original parameters of the potential (no summation convention)
\begin{equation} \label{m18}
(\mu^2_1 + \mu^2_2)\kappa_i^{*}\kappa'^*_i=\lambda^*_6 \kappa_i^{*2}+
\lambda_6 \kappa'^{*2}_i\, .
\end{equation}
Analogous to our $\not\!\!R$ model is the limiting case of the MSSM,
without explicit $R$-parity breaking, if one sets $\kappa_i=\kappa'_i=0$
and replaces $\mu^2_{L_{ij}}$ by the diagonal coupling $\mu^2_{L_i}$.
Again, the attempt to maintain $w_i \neq 0$ yields
the `fine-tuning' relation (see also \cite{comel}).
\begin{equation} \label{m19}
(\mu_1^2 - \mu^2_{L_i})(\mu^2_2 + \mu^2_{L_i})=\vert \lambda_6 \vert^2\, .
\end{equation}
This demonstrates nicely how, unlike the MSSM, the $R$-parity-breaking
couplings $\varepsilon_i$ from Eq.~(\ref{m3}) naturally lead to non-zero
vev's for the scalar neutrinos. Had we neglected the bilinear term
$\varepsilon_{ab} \varepsilon_i \hat{L}_i^a \hat{H}_2^b$ in Eq.~(\ref{m3}), as
done
usually, then there would have been no compelling reason to acquire non-zero
vev's for the scalar neutrino fields even in the $R$-parity broken case.
Moreover, if we conserve an individual lepton number, say $L_i$, in
the Lagrangian/superpotential of the $R$-parity broken SUSY model ---among
other couplings, this implies that $\varepsilon_i=0$---, this symmetry will
not break spontaneously; so, we are free to choose the corresponding
$\varphi_i$ to have a vanishing vev.
This follows again from Eqs.\ (\ref{m18}) and (\ref{m19}).

The spontaneous breaking of CP violation in the case of $V_{Scalar}$,
Eq.~(\ref{m9}), has been discussed in~\cite{JNCP}. We supplement these
considerations on CP properties by giving below all conditions necessary to
restore CP conservation in $V^{2H}+V^L+V^{\not L}$.
Denoting $\eta_{{}_1}$, $\eta_{{}_2}$
and $\eta_{{}_{L_i}}$ the CP phases of the fields
$\phi_1$, $\phi_2$, and $\varphi_i$, respectively, we find from the
requirement of CP conservation that (no summation convention below)
\begin{eqnarray} \label{m20}
&&\lambda_6^* \eta_{{}_1} \eta_{{}_2}^*=\lambda_6\, , \nonumber \\
&&\kappa_{jk}=\kappa^*_{jk}\eta_{{}_{L_i}}\eta^*_{{}_{L_k}}\, , \nonumber \\
&&\kappa_{nmjk}=\kappa^*_{nmjk}\eta_{{}_{L_n}}
\eta_{{}_{L_m}}\eta^*_{{}_{L_i}}\eta^*_{{}_{L_j}}\, ,
\nonumber \\
&&\kappa^*_i \eta^*_{{}_1} \eta^*_{{}_{L_i}}=\kappa_i\, , \nonumber \\
&&\kappa'^*_j \eta^*_{{}_2} \eta^*_{{}_{L_j}}=\kappa'_j\, ,  \nonumber \\
&&\kappa^*_{nmj}\eta_{{}_1} \eta^*_{{}_{L_j}}
\eta_{{}_{L_n}}\eta_{{}_{L_m}}=\kappa_{nmj}\, .
\end{eqnarray}
{}From Eq.\ (\ref{m20}), one can derive over twenty conditions for
CP conservation which in contrast
to (\ref{m20}) do not involve the CP phases $\eta$'s. However, it is obvious
that not all such conditions are independent. For instance, a set of
independent conditions that follows from (\ref{m20}) is given by
\begin{eqnarray} \label{m21}
&&\Im m(\lambda_6 \kappa_i \kappa'^*_j \kappa_{ij})=0\, , \nonumber \\
&&\Im m(\kappa_i \kappa^*_j \kappa_{ij})=0\, , \nonumber \\
&&\Im m(\kappa'_i \kappa'^*_j \kappa_{ij})=0\, , \nonumber \\
&&\Im m(\kappa^*_{ni}\kappa^*_{mj}\kappa_{nmij})=0\, , \nonumber \\
&&\Im m(\kappa^*_{n}\kappa_{mj}\kappa^*_{nmj})=0\, ,
\end{eqnarray}
where again no summation convention has been used.
After spontaneous symmetry breaking, the first equation in the
set (\ref{m21}) for $i=j$, {\em i.e.}\ $\Im m(\lambda_6 \kappa_i
\kappa^{'*}_i)=0$, translates  into the following three equivalent
conditions:
\begin{eqnarray} \label{m22}
&&\Im m(\kappa_i w_i v_1)=0\, , \nonumber \\
&&\Im m(\kappa'_i w_i v_2)=0\, , \nonumber \\
&&\Im m(\lambda_6 v_1^* v_2)=0\, .
\end{eqnarray}
For the case at hand, it is, however, possible to have
complex vev's such that CP gets broken spontaneously.
For further details on this issue, the reader is referred to~\cite{JNCP}.

Some remarks on the vev's of sneutrinos, $w_i$, are in order.
One should notice that through the kinetic term
\begin{displaymath} 
\sum_i \left(D_{\mu}{\cal S}_i
\right)^{\dagger}\left(D^{\mu}{\cal S}_i\right)\, ,
\end{displaymath}
where
$D_{\mu}$ is the covariant derivative and ${\cal S}_i$ are the scalar fields in
the theory, the sneutrino vev's contribute to the gauge bosons masses. In this
way, the SM vev is obtained by
\begin{equation} \label{mex2}
v \equiv \sqrt{v_1^2 + v_2^2 + \sum_i w_i^2}\ =\ \frac{2M_W}{g}\, .
\end{equation}
As a consequence, $w_i$ and the angle $\beta$ defined by
\begin{equation} \label{mex3}
\tan \beta = { v_1 \over v_2}\, ,
\end{equation}
for real $v_i$, may be regarded as free parameters of the theory,
while $v_i$ are not free any longer, but determined by
\begin{eqnarray} \label{mex4}
&&v_1 = \sin \beta \sqrt{v^2 -\sum_i w^2_i}\, , \nonumber \\
&&v_2 = \cos \beta \sqrt{v^2 -\sum_i w^2_i}\, .
\end{eqnarray}
Evidently, the vev's of the scalar neutrinos, $w_i$, cannot have arbitrarily
large values, but they are bounded from above,
as can be readily seen from Eqs.~(\ref{mex2}) and (\ref{mex4}).

\setcounter{equation}{0}
\section{Mass matrices}
\indent

In this section, we will present the mass matrices of the neutralino/neutrino
as well as those
of the chargino/charged lepton states. Since the lepton number is
explicitly broken, the neutrinos will mix with the neutralinos to give the
neutrinos mass. A natural seesaw mechanism emerges, in which
$\mu$, $v_1$, $v_2$, and the gaugino mass parameters $M$ and $M'$
act as the heavy scales, and the lepton-number-breaking couplings
$\varepsilon_i$
together with the sneutrino vev's $w_i$ constitute the light Dirac components
of the seesaw matrix.
It will turn out that only one neutrino becomes
massive through this mechanism at the tree level.
These considerations are relevant for putting limits on the
lepton-number-breaking parameters. In particular, one can already infer
constraints on those parameters from the $\tau$-neutrino mass.
Furthermore, the neutralino--neutrino or chargino--charged lepton mixing
will enter the interaction Lagrangians of $W$ and $Z$, giving rise
to non-SM processes, through which the new parameters can also be constrained.

\subsection{Neutralino--neutrino mixing}
\indent

In general, there are two mechanisms that can give rise to neutrino masses
in the Born approximation. For example, one possibility is to give masses
to the left-handed neutrinos through the vev of an exotic Higgs field which
transform under $SU(2)_L$ as a triplet. The other mechanism requires,
in general, the mixing of the left-handed neutrinos with other neutral
fields of the theory. The latter are usually taken to be the
right-handed neutrinos, introducing hereby additional fields in the theory. In
our minimal $R$-parity broken SUSY model, in which right-handed neutrinos
are absent, the r\^ole of the new neutral fields required for the
afore-mentioned mixing will be assumed by the gauginos and higgsinos.

In two component notation, let $\Psi'$ denote the column vector of neutrinos
and
neutralinos
\begin{equation} \label{m23}
\Psi^{'T}_0=(\psi^1_{L_1},\,\, \psi^1_{L_2},\,\, \psi^1_{L_3},\,\,
-i\lambda',\,\, -i\lambda_3,\,\, \psi^1_{H_1},\,\, \psi^2_{H_2})\, ,
\end{equation}
where $\psi^1_{L_i}$ are the neutrino fields ---the upper index indicates
the component of the doublet---, $-i\lambda'$ and $-i\lambda_3$ are the
unmixed photino and gaugino states, respectively, and the last two entries
refer to the two higgsino fields. In the Weyl basis, the Lagrangian
describing the neutralino/neutrino masses is then given by
\begin{equation} \label{m24}
{\cal L}^{\chi^0}_{mass}=-{1 \over 2}\Psi_0^{'T}{\cal M}_0 \Psi'_0\ +\
\mbox{H.c.}\, ,
\end{equation}
where the mass matrix has the general seesaw-type structure
\begin{equation} \label{m25}
{\cal M}_0=\left(\begin{array}{cc}
0 & m \\
m^T & M_4 \end{array}\right) .
\end{equation}
Here, the sub-matrix $m$ is the following $3 \times 4$ dimensional matrix:
\begin{equation} \label{m26}
m=\left(\begin{array}{cccc}
-{1 \over 2}g' w_e    & {1 \over 2}g w_e    & 0 & -\varepsilon_e \\
-{1 \over 2}g' w_\mu  & {1 \over 2}g w_\mu  & 0 & -\varepsilon_\mu \\
-{1 \over 2}g' w_\tau & {1 \over 2}g w_\tau & 0 & -\varepsilon_\tau
\end{array}\right) .
\end{equation}
In Eq.~(\ref{m25}), $M_4$ is the usual $4 \times 4$ dimensional
neutralino mass matrix of the MSSM, which has the form
\begin{equation} \label{m27}
M_4=\left(\begin{array}{cccc}
cM & 0 & -{1 \over 2}g'v_1 & {1 \over 2}g'v_2 \\
0 & M & {1 \over 2}gv_1 & -{1 \over 2}gv_2 \\
-{1 \over 2}g'v_1 & {1 \over 2}gv_1 & 0 & - \mu \\
{1 \over 2}g'v_2 & -{1 \over 2}gv_2 & - \mu & 0 \end{array}\right) ,
\end{equation}
where $M$ is the common gaugino mass parameter and
$\displaystyle{c={5 g'^2 \over 3g^2 }\simeq 0.5}$.
The seesaw hierarchy is now evident, when constraints on neutralino masses
and upper limits on lepton-number-violating couplings will be considered
in Section~5. We will then find that $(M_4)_{ij} \gg m_{kl}$ in agreement
with experimental constraints on neutralino and neutrino masses.
We can utilize this posterior fact to calculate the diagonalization
of ${\cal M}_0$ in an approximate way in terms of the small matrix-valued
quantity defined as
\begin{equation}\label{xi}
\xi=mM_4^{-1}\, .
\end{equation}
Parenthetically, we wish to draw the reader's attention to one
exact result in connection with the diagonalization of ${\cal M}_0$.
Because of the different hypercharge assignments of the two higgsinos
and the absence of light-neutrino masses at the tree level,
the first three lines together with the last line in ${\cal M}_0$ are not
linearly independent. As an immediate consequence of the latter, two neutrino
masses are exactly zero in the Born approximation \cite{JN}.

Let us now define the mass eigenstates $\Psi_0$ by the rotation
\begin{eqnarray} \label{m28}
\Psi_{0i}=\Xi_{ij} \Psi'_{0j}\, , \nonumber \\
\Xi^*{\cal M}_0 \Xi^{\dagger} =\widehat{{\cal M}}_0\, ,
\end{eqnarray}
where $\widehat{{\cal M}}_0$ is the diagonal matrix with neutrino/neutralino
masses as elements. To leading order in $\xi$ expansion,
the approximate form of $\Xi^*$ is readily estimated to be
\begin{equation} \label{m30}
\Xi^*=\left(\begin{array}{cc}
V_\nu^T & 0 \\
0 & N^* \end{array}\right)
\left(\begin{array}{cc}
1 -{1 \over 2}\xi \xi^{\dagger} & -\xi \\
\xi^{\dagger} &  1 -{1 \over 2}\xi^\dagger \xi
\end{array}\right),
\end{equation}
where the second matrix block-diagonalizes ${\cal M}_0$ to the form
$\mbox{diag}(m_{eff}, M_4)$ with
\begin{equation} \label{m31}
m_{eff}=-m\; M_4^{-1}\; m^T = {cg^2+ g'^2 \over D}\,
\left(\begin{array}{ccc}
\Lambda_e^2 & \Lambda_e \Lambda_\mu
& \Lambda_e \Lambda_\tau \\
\Lambda_e \Lambda_\mu & \Lambda_\mu^2
& \Lambda_\mu \Lambda_\tau \\
\Lambda_e \Lambda_\tau & \Lambda_\mu \Lambda_\tau & \Lambda_\tau^2
\end{array}\right).
\end{equation}
The quantities $\Lambda_i$ and $D$ newly introduced are defined as follows:
\begin{equation} \label{31}
\vec{\Lambda}\equiv \mu \vec{w} - v_1 \vec{\varepsilon}\, ,
\end{equation}
and
\begin{equation} \label{32}
D \equiv 4{det M_4 \over M}=2\mu \left[-2cM\mu + v_1v_2
\left(cg^2 + g'^2 \right)\right] .
\end{equation}
The sub-matrices $N$ and $V_\nu$ in Eq.~(\ref{m30}) diagonalize $M_4$ and
$m_{eff}$ in the following way:
\begin{equation} \label{33}
N^* M_4 N^{\dagger}=\mbox{diag}(m_{\tilde{\chi}^0_i})\, ,
\end{equation}
where $m_{\tilde{\chi}^0_i}$ are the heavy neutralino masses only.
For the diagonalization of $M_4$, we have retained the notation and
convention of Ref.~\cite{HK}.
For the neutrino case, we obtain
\begin{equation}
\label{m34}
V^T_{\nu}\; m_{eff}\; V_{\nu} = \mbox{diag}(0, \; 0, \; m_{\nu}) \, ,
\end{equation}
where the only non-zero neutrino mass is given by
\begin{equation} \label{m35}
m_{\nu}=tr(m_{eff})={cg^2+ g'^2 \over D}\, \vert \vec{\Lambda} \vert^2\, .
\end{equation}
Furthermore, an analytic calculation of the rotation matrix $V_{\nu}$
gives \cite{JN}
\begin{equation} \label{m36}
V_{\nu}=\left(\begin{array}{ccc}
\cos \theta_{13} & 0 & -\sin \theta_{13} \\
\sin \theta_{23}\sin \theta_{13} & \cos \theta_{23} & \sin \theta_{23}
\cos \theta_{13} \\
\sin \theta_{13} & \sin \theta_{23} & \cos \theta_{13}\cos \theta_{23}
\end{array}\right) ,
\end{equation}
where the mixing angles are expressed through the vector
$\vec{\Lambda}$ as follows:
\begin{equation} \label{m37}
\tan \theta_{13} = -{\Lambda_e \over
\sqrt{\Lambda_\mu^2 + \Lambda_\tau^2}}, \;\;\;\;\;
\tan \theta_{23} = {\Lambda_\mu \over \Lambda_\tau} \, .
\end{equation}
In~\cite{JN}, the baryogenesis constraint on all lepton-number-violating
couplings were applied, which led to a solution to the solar neutrino puzzle
through vacuum oscillations.
In that case, the neutrino mass $m_\nu$ came out rather naturally of order
$10^{-5}$ eV, while the mixing angle $\theta_{13}$ was predicted to be
large, {\em i.e.}\ $\tan\theta_{13} \simeq -1/\sqrt{2}$.
Since we can evade the constraints from BAU by conserving one individual lepton
number, our scenario regarding the light neutrino mass, $m_\nu$,
is quite different.

In Section 6, we will discuss some numerical examples of the neutrino mass
as well as the resulting constraints on $w_i$, $\varepsilon_i$ together with
the constraints emerging from exotic processes.
Here we note in passing
that appreciable values for $w_i$ and $\varepsilon_i$ in the GeV range
result in a tau-neutrino mass of ${\cal O}($MeV) which is still allowed
by laboratory constraints. Let us now demonstrate explicitly, by an
example, how the $\varepsilon_i$ terms can change
some of the phenomenological implications. We choose the following set
of parameters: $M=\mu=2M_W$, $\tan\beta =1$, $\varepsilon_{\tau}=w_{\tau}=0$,
$w_e=w_{\mu}\equiv w=1$~GeV, $\varepsilon_{e}=\varepsilon_{\mu} \equiv
\varepsilon$. If we now put $\varepsilon=0$, then $m_{\nu} \simeq 3.5$~MeV.
On the other hand, the same soft-SUSY parameters and vev's,
but having now $\varepsilon=4w$ instead, give $m_{\nu} \simeq 38.5$~MeV,
which already exceeds the laboratory limit on the tau-neutrino mass.

\subsection{Chargino--charged lepton mixing}
\indent

Similar to the case of neutralino--neutrino mixing, the explicit violation
of the lepton number allows also for chargino--charged lepton mixing.
In two component notation, the mass term takes the form
\begin{equation} \label{m38}
{\cal L}^{\chi^+}_{mass}= -\zeta^{'T} {\cal M}_+\, \omega'\ +\ \mbox{H.c.}\, ,
\end{equation}
where in the vector $\zeta'$, we gather the lower components of a charged
Dirac spinor in the Weyl representation, {\em i.e.}\
\begin{equation} \label{m39}
\zeta^{'T}=(\psi^2_{L_1},\,\, \psi^2_{L_2},\,\, \psi^2_{L_3},\,\,
-i\lambda_-,\,\,  \psi^2_{H_1})\, ,
\end{equation}
whereas $\omega'$ contains the upper components
\begin{equation} \label{m40}
\omega^{'T}=(\psi_{R_1},\,\, \psi_{R_2},\,\, \psi_{R_3},\,\,
-i\lambda_+,\,\,  \psi^1_{H_2})\, .
\end{equation}
In order to establish contact between the notation of the MSSM in~\cite{HK}
or that of our minimal $\not\!\! R$ model and the usual SM notation, we
note that the charged leptons are represented
by their charged conjugate fields, {\em i.e.}
\begin{displaymath}
l_i^C\ =\ \left( \begin{array}{c}
\psi_{R_i} \\ \bar{\psi}_{L_i}^2 \end{array} \right).
\end{displaymath}
In this basis, the chargino/charged-lepton mass matrix ${\cal M}_+$
appearing in Eq.~(\ref{m38}) may be written down as
\begin{equation} \label{m41}
{\cal M}_+=\left(\begin{array}{cc}
M_l & E \\
E' & S
\end{array}\right) ,
\end{equation}
where $S$ is the usual MSSM chargino mass matrix given by
\begin{equation} \label{m42}
S=\left(\begin{array}{cc}
M & \frac{1}{\sqrt{2}} g v_2 \\
\frac{1}{\sqrt{2}} g v_1 & \mu
\end{array}\right) .
\end{equation}
The sub-matrices $E$ and $E'$ which give rise to chargino--charged lepton
mixing
are defined as follows:
\begin{equation} \label{m43}
\displaystyle{E=\left(\begin{array}{cc}
{1\over \sqrt{2}} g w_e & \varepsilon_e \\
{1\over \sqrt{2}} g w_\mu & \varepsilon_{\mu} \\
{1\over \sqrt{2}} g w_\tau & \varepsilon_{\tau}
\end{array}\right)} ,
\end{equation}
and
\begin{equation} \label{m44}
E'=\left(\begin{array}{ccc}
0 & 0 & 0 \\
\Upsilon_e & \Upsilon_{\mu} & \Upsilon_{\tau}
\end{array}\right) ,
\end{equation}
where $\displaystyle{\Upsilon_l \sim {m_l \over v_1} w_l}$ and
$m_l$ are the lepton masses. For our numerical purposes,
we will assume that $M_l$ is a diagonal matrix whose elements can be
identified, to a high accuracy, with the physical lepton masses $m_i$.
In addition, we can neglect the elements of $E'$ as
compared to the other entries in Eq.~(\ref{m41}). Therefore, we will be
working in the approximation $E'=0$.

Let us now express the mass eigenstates $\zeta$ and $\omega$ in terms
of the states $\zeta'$ and $\omega'$ via the unitary transformations
\begin{equation} \label{m45}
\zeta_i=\Sigma_{ij}\zeta'_j, \,\,\, \omega=\Omega_{ij}\omega'_j\, .
\end{equation}
The bi-diagonalization leads then to the diagonal matrix $\widehat{{\cal M}}_+$
whose elements are the chargino and lepton masses
\begin{equation} \label{m46}
\Sigma^* {\cal M}_+ \Omega^{\dagger}=\widehat{{\cal M}}_+\, .
\end{equation}
Proceeding now as in the case of the neutralino--neutrino mixing, we
carry out an approximate diagonalization for ${\cal M}_+$.
In this way, the expansion parameters are found to be
\begin{eqnarray} \label{m47}
&&\xi_{{}_{L}}^*=ES^{-1} \, ,\nonumber \\
&&\xi_{{}_{R}}^*=M_l^{\dagger}ES^{-1}(S^{-1})^T
=M_l^{\dagger}\xi_{{}_{L}}^*(S^{-1})^T\, .
\end{eqnarray}
Note that $\xi_{{}_{R}} \sim \xi_{{}_{L}} m_l/M$.
To leading order in $\xi_{{}_L}$ and $\xi_{{}_R}$, the rotation matrices
are written down as
\begin{equation} \label{m48}
\Sigma^*=\left(\begin{array}{cc}
V_L & 0 \\
0 & U^* \end{array}\right)\left(\begin{array}{cc}
1 -{1 \over 2} \xi_{{}_{L}}^*\xi_{{}_{L}}^T & -\xi_{{}_{L}}^* \\
\xi_{{}_{L}}^T & 1 -{1 \over 2} \xi_{{}_{L}}^T\xi_{{}_{L}}^*
\end{array}\right) ,
\end{equation}
and
\begin{equation} \label{m49}
\Omega^{\dagger}=
\left(\begin{array}{cc}
1 -{1 \over 2} \xi_{{}_{R}}^*\xi_{{}_{R}}^T & \xi_{{}_{R}}^* \\
-\xi_{{}_{R}}^T & 1 -{1 \over 2} \xi_{{}_{R}}^T\xi_{{}_{R}}^*
\end{array}\right)
\left(\begin{array}{cc}
V_R^{\dagger} & 0 \\
0 & V^{\dagger} \end{array}\right).
\end{equation}
Adopting the convention of~\cite{HK} for the matrices, which
also appear in the MSSM, we have
\begin{eqnarray} \label{m50}
&& U^*SV^{\dagger}=\widehat{S}\, , \nonumber \\
&& V_L M_l V_R^{\dagger} =\widehat{M}_l\, ,
\end{eqnarray}
where, as before, the hatted matrices are diagonal.

\setcounter{equation}{0}
\section{The $W$- and $Z$-boson interaction Lagrangians}
\indent

In this section, we will derive the interaction Lagrangians of $Z$ and $W$
bosons with neutralinos/neutrinos and charginos/charged leptons. We will
first present general expressions and, subsequently, use the analytic
results of the approximate diagonalization of the mass matrices, given in
the preceeding section, to calculate the mixing matrices in the first
order approximation. Here and in the following,
because of the mixing,  we collectively call $\Psi_0$ all the neutralinos,
with neutrinos being the light neutralinos, and $\zeta$, $\omega$
all the charginos, where the charged leptons are the light charginos.

\subsection{General expressions}
\indent

Starting from two component notation and defining for convenience the matrix
\begin{equation} \label{m51}
T^Z=\mbox{diag}(1,\, 1,\, 1,\, 0,\, 0,\, 1,\, -1)\, ,
\end{equation}
the interaction Lagrangian of $Z$ with neutralinos reads
\begin{equation} \label{m52}
{\cal L}^{Z\chi^0\chi^0}_{int}\ =\ -{g \over 2\cos\theta_w}
Z^{\mu}\bar{\Psi}'_{0i} T^Z_{ij}\bar{\sigma}_{\mu}\Psi'_{0j} \, .
\end{equation}
After replacing the weak eigenstates $\Psi'_{0i}$ by four component
Majorana mass eigenstates $\chi^0_i$ in Eq.~(\ref{m52}), we obtain
\begin{equation} \label{Znunu}
{\cal L}^{Z\chi^0\chi^0}_{int}\ =\ - \frac{g}{4\cos\theta_w}\, Z^\mu\,
\bar{\chi}^0_i \gamma_\mu\Big( i \Im m \tilde{C}_{ij}\
-\ \gamma_5 \Re e \tilde{C}_{ij}\Big) \chi^0_j\, ,
\end{equation}
where
\begin{equation} \label{m54}
\tilde{C}_{ij}\ =\ (\tilde{C}^{\dagger})_{ij}\ =\ (\Xi T^Z \Xi^{\dagger})_{ij}
\, .
\end{equation}
It is easy to check that Eq.~(\ref{Znunu}) reproduces the
$Z$-neutralino-neutralino interaction of the MSSM, when the leptonic
$\not\!\! R$ admixture is neglected.

To calculate the $Z$-chargino-chargino coupling,  we again define two
auxiliary matrices
\begin{eqnarray} \label{m55}
T^Z_R &=& \mbox{diag}(0,\,0,\, 0,\,2,\, 1)\, , \nonumber \\
T^Z_L &=& \mbox{diag}(1,\,1,\, 1,\,2,\, 1)\, ,
\end{eqnarray}
such that ${\cal L}^{Z\bar{\chi}^-\chi^-}_{int}$ takes the form
\begin{eqnarray} \label{m56}
{\cal L}^{Z\bar{\chi}^-\chi^-}_{int}
&=&{g \over 2\cos \theta_w}Z^{\mu}\Big[
\bar{\zeta}'_i(T^Z_L)_{ij}\bar{\sigma}_{\mu}\zeta'_j
- \bar{\omega}'_i(T^Z_R)_{ij}\bar{\sigma}_{\mu}\omega'_j
\nonumber \\
&& +2 \sin^2 \theta_w ( \bar{\omega}'_i
\bar{\sigma}_{\mu}\omega'_i
-\bar{\zeta}'_i\bar{\sigma}_{\mu}\zeta'_i )\Big] .
\end{eqnarray}
Denoting by $\chi^{-}_i$ the physical charginos in the
four-component Dirac notation pertaining to the definition of Eq.~(\ref{m39}),
the above Lagrangian can be written down as
\begin{equation}\label{Zll}
{\cal L}^{Z\bar{\chi}^-\chi^-}_{int}\ =\ \frac{g}{2\cos\theta_w}\, Z^\mu\,
\bar{\chi}^-_i \gamma_\mu \Big( \tilde{A}^L_{ij} \mbox{P}_L\
+\ \tilde{A}^R_{ij} \mbox{P}_R\Big)\chi^-_j\, ,\\
\end{equation}
where P$_L$(P$_R)=[1-(+)\gamma_5]/2$ and
\begin{eqnarray} \label{m58}
\tilde{A}^L_{ij} &=& (\Sigma T^Z_L \Sigma^{\dagger})_{ij} - 2\delta_{ij}
\sin^2 \theta_w\, , \nonumber \\
\tilde{A}^R_{ij}&=&(\Omega^* T^Z_R \Omega^T)_{ij} - 2\delta_{ij}
\sin^2 \theta_w \, .
\end{eqnarray}
As done above for the neutral-current interactions, for the charged-current
case we first introduce two auxiliary matrices given by
\begin{equation} \label{m59}
T^L=\left(\begin{array}{ccccccc}
1 & 0 & 0 & 0 & 0 & 0 & 0 \\
0 & 1 & 0 & 0 & 0 & 0 & 0 \\
0 & 0 & 1 & 0 & 0 & 0 & 0 \\
0 & 0 & 0 & 0 & \sqrt{2} & 0 & 0 \\
0 & 0 & 0 & 0 & 0 & 1 & 0 \end{array}\right), \, \, \,
T^R=\left(\begin{array}{ccccc}
0 & 0 & 0 & 0 & 0 \\
0 & 0 & 0 & 0 & 0 \\
0 & 0 & 0 & 0 & 0 \\
0 & 0 & 0 & 0 & 0 \\
0 & 0 & 0 & -\sqrt{2} & 0 \\
0 & 0 & 0 & 0 & 0 \\
0 & 0 & 0 & 0 & 1 \end{array}\right) .
\end{equation}
In the Weyl weak eigenbasis, we can then write
\begin{equation} \label{m60}
{\cal L}^{W\chi^-\chi^0}_{int}\ =\ -{ g \over \sqrt{2}} \Big(
\bar{\zeta}'_i T^L_{ij}\bar{\sigma}^{\mu}\Psi'_{0j}  +
\bar{\Psi'}_{0i} T^R_{ij}\bar{\sigma}^{\mu}\omega'_j\Big) W^-_{\mu}\
+\ \mbox{H.c.}\, ,
\end{equation}
and
\begin{equation}\label{Wlnu}
{\cal L}^{W\chi^-\chi^0}_{int} \ =\ -\frac{g}{\sqrt{2}}\, W^{-\mu}\,
\bar{\chi}^-_i \gamma_\mu \Big( \tilde{B}^L_{ij} \mbox{P}_L\
+\ \tilde{B}^R_{ij} \mbox{P}_R\Big) \chi^0_j\ +\ \mbox{H.c.}\, ,\\
\end{equation}
in four-component mass eigenbasis notation. The mixing matrices are
\begin{eqnarray} \label{m62}
\tilde{B}^L_{ij} &=& (\Sigma T^L \Xi^\dagger)_{ij}\, , \nonumber \\
\tilde{B}^R_{ij} &=& -[\Omega^* (T^R)^T \Xi^T ]_{ij}\, .
\end{eqnarray}
In the next section, we will give analytic approximate expressions for
all mixing matrices defined above.

\subsection{Mixing matrices}
\indent

In Section~4.1, we have derived the analytic expressions of
the interaction Lagrangians of the $W$ and $Z$ bosons with chargino and
neutralino states in the $R$-parity-violating SUSY model.
However, the mixings $\tilde{A}^L$, $\tilde{A}^R$,
$\tilde{B}^L$, $\tilde{B}^R$, and $\tilde{C}$ that govern
these interactions are high dimensional matrices, involving
a large number of parameters. Therefore, it is more convenient
to find approximative forms for the mixing matrices that will
enable us to appreciate the strength of the $Z$- and $W$-boson
couplings in the model under consideration.

To facilitate our presentation, we first introduce the following
auxiliary matrices:
\begin{eqnarray}
d &=& \mbox{diag}(2,1)\, ,\qquad t_z\ =\ \mbox{diag}(0,0,1,-1)\, ,\nonumber\\
t_{{}_L}& =& \left( \begin{array}{cccc}
0 & \sqrt{2} & 0 & 0 \\
0 & 0 & 1 & 0 \end{array} \right),\nonumber\\
t_{{}_R}& =& \left( \begin{array}{cccc}
0 & \sqrt{2} & 0 & 0 \\
0 & 0 & 0 & -1 \end{array} \right).
\end{eqnarray}
Substituting the unitary mixing matrices of Eqs.\ (\ref{m30}), (\ref{m48}),
and (\ref{m49}) into Eqs.~(\ref{m54}), (\ref{m58}), and (\ref{m62}),
and neglecting terms of ${\cal O}(\xi_{{}_L}^3,\xi^3 )$ and higher,
we obtain
\begin{eqnarray}
\tilde{A}^L &=& -2\sin^2\theta_w {\mbox{\bf 1}}\ +\ \left(
\begin{array}{cc}
1-\tilde{\xi}_{{}_L}(1-d)\tilde{\xi}_{{}_L}^\dagger &
\tilde{\xi}_{{}_L}(1-d)U^\dagger\\
U(1-d)\tilde{\xi}^\dagger_{{}_L} & UdU^\dagger \end{array}\right), \label{AL}\\
\tilde{A}^R &=& -2\sin^2\theta_w {\mbox{\bf 1}}\ +\ \left(
\begin{array}{cc}
0 & -\tilde{\xi}_{{}_R}dV^T\\
-V^*d\tilde{\xi}_{{}_R}^\dagger & V^*d V^T \end{array}\right), \label{AR}\\
\tilde{B}^L &=& \left( \begin{array}{cc}
V^l-\frac{1}{2}\tilde{\xi}_{{}_L}\tilde{\xi}^\dagger_{{}_L}V^l-
\frac{1}{2}V^l\tilde{\xi}^*\tilde{\xi}^T +
\tilde{\xi}_{{}_L} t_{{}_L} \tilde{\xi}^T
 & (V^l\tilde{\xi}^* - \tilde{\xi}_{{}_L} t_{{}_L}) N^\dagger \\
U(\tilde{\xi}^\dagger_{{}_L}V^l-t_{{}_L}\tilde{\xi}^T) & Ut_{{}_L}N^\dagger
\end{array} \right),\label{BL}\\
\tilde{B}^R &=& \left( \begin{array}{cc}
0 & - \tilde{\xi}_{{}_R} t_{{}_R} N^T \\
-V^* t_{{}_R} \tilde{\xi}^\dagger & V^* t_{{}_R} N^T \end{array}
\right),\label{BR}\\
\tilde{C} &=& \left( \begin{array}{cc}
1- \tilde{\xi}^* (1-t_z)\tilde{\xi}^T & \tilde{\xi}^* (1-t_z)  N^\dagger \\
N(1-t_z)\tilde{\xi}^T & N t_z N^\dagger \end{array} \right).\label{C}
\end{eqnarray}
Here, we have defined $\tilde{\xi}_{{}_L}
=V_L^*\xi_{{}_L}$,
$\tilde{\xi}_{{}_R}=\widehat{M}_l \tilde{\xi}_{{}_L} (S^{-1})^\dagger$,
$\tilde{\xi} = V_\nu^T \xi$, and $V^l=V_L^* V_\nu$.
Furthermore, the unitary matrices $V_L$, $V_R$, $V_\nu$,
$U$, $V$, $N$, together with the mixing matrices $\xi_{{}_L}$,
$\xi_{{}_R}$, and $\xi$ are defined in Section~3. In the derivation
of Eqs.~(\ref{AR}) and (\ref{BR}), we have also used the fact that
$\xi_{{}_R} ={\cal O}( \xi_{{}_L} m_l/M)$.

{}From Eqs.~(\ref{AL})--(\ref{C}), it is now easy to see how the
$R$-parity-violating couplings to ordinary leptons deviate from the
SM vertices. To leading order in $\xi_{{}_L}$ and $\xi$, we find that the
interactions of the $W$ and $Z$ bosons with left-handed charged leptons
and neutrinos are modified, whereas the corresponding couplings to
right-handed charged leptons remain unaffected, having
the SM form.

\setcounter{equation}{0}
\section{Laboratory and cosmological constraints}
\indent

Our aim is to constrain the parameter space of this
$\not \!\! R$ scenario, by taking laboratory and
cosmological constraints into account. For this purpose,
we will pay special attention to limits derived from
low-energy processes and LEP data, such as charged lepton decays
of the form $l^-\to l'^-l_1^-l_1^+$, flavour-changing
$Z$-boson decays $Z\to l_il_j$, the invisible width
of the $Z$ boson, charged-current universality in muon
and tau decays, lepton universality at the $Z$ peak,
and charged-current universality in pion decays.
In this vein, we will report some phenomenological implications
of our minimal model that may be relevant to explain the intriguing
anomaly found by the KARMEN collaboration~\cite{KARMEN}. In the last section,
we will discuss the viability of our model when cosmological constraints are
considered, such as the requirement of not washing out the primordial BAU and
the absence of large disruptive reheating effects caused by an unstable $\tau$
neutrino with $m_{\nu_\tau}={\cal O}(10)$~MeV.

\subsection{{\boldmath $l^-\to l'^-l_1^-l^+_1$}}
\indent

As has been found in Lagrangian~(\ref{Zll}), the model
predicts flavour-changing neutral current (FCNC) $Zll'$ couplings
at the tree level. These new $\not\!\! R$ interactions induce
$\tau$ and $\mu$ decays into three lighter charged leptons. In
this way, we obtain
\begin{equation}
B(l^-\to l'^-l_1^-l^+_1)\ =\
\frac{\alpha_w^2 m^4_l}{1536\pi M_W^4}\, \frac{m_l}{\Gamma_l}
\left( \left|\tilde{A}^L_{ll'}\right|^2
+\left|\tilde{A}^R_{ll'}\right|^2\right)\left(\left|\tilde{A}^L_{l_1l_1}
\right|^2 +\left|\tilde{A}^R_{l_1l_1}\right|^2\right), \label{BRlll}
\end{equation}
where $\alpha_w=g^2/4\pi$ and $\Gamma_l$ is the total width of the decaying
charged lepton $l$.

The experimental upper limit on the branching ratio of $\mu^- \to e^-e^-e^+$
is given by~\cite{PDG94}
\begin{equation}
B(\mu^- \to e^-e^-e^+) \ \leq\ 1.0\cdot 10^{-12}\, ,\label{Bexpmueee}
\end{equation}
at $90\%$ confidence level (CL).
Recently, CLEO collaboration~\cite{CLEO} has considerably lowered
experimental upper bounds on branching ratios of
neutrinoless $\tau$-lepton decays. They have found
\begin{eqnarray}
B(\tau^-\to e^- e^+ e^- )      &\leq& 3.3\cdot 10^{-6}\, ,\nonumber\\
B(\tau^-\to \mu^- e^+ e^- )    &\leq& 3.4\cdot 10^{-6}\, ,\nonumber\\
B(\tau^-\to e^- \mu^+ \mu^- )  &\leq& 3.6\cdot 10^{-6}\, ,\nonumber\\
B(\tau^-\to \mu^- \mu^+ \mu^- )&\leq& 4.3\cdot 10^{-6}\, ,\label{Bexptaulll}
\end{eqnarray}
at $90\%$ CL. Theoretical predictions obtained for the observables
given in Eqs.~(\ref{Bexpmueee}) and~(\ref{Bexptaulll}) will be discussed
in Section 6.

\subsection{{\boldmath $Z\to l^-l'^+$} and {\boldmath $Z\to \nu\nu$}}
\indent

The presence of FCNC $Zll'$ couplings at the tree level will also
give rise to flavour-violating $Z$-boson decays at LEP.  The theoretical
prediction of their branching ratios is determined by
\begin{equation}
B(Z\to l^-l'^+\ \mbox{or}\
l^+ l'^-)\ =\ \frac{\alpha_w}{12\cos^2\theta_w}\,
\frac{M_Z}{\Gamma_Z}\left( \left|\tilde{A}^L_{ll'}\right|^2
+\left|\tilde{A}^R_{ll'}\right|^2 \right),\label{BZlilj}
\end{equation}
where $\Gamma_Z=2.49$~GeV is the total width of the $Z$ boson
measured experimentally~\cite{PDG94}. Furthermore, an analysis
of this kind of decays at LEP yields
\begin{eqnarray}
B( Z\to e^-\mu^+\ \mbox{or}\ e^+\mu^-)   &\leq & 6.0\cdot 10^{-6}\,
,\nonumber\\
B( Z\to e^-\tau^+\ \mbox{or}\ e^+\tau^-) &\leq & 1.3\cdot 10^{-5}\,
,\nonumber\\
B( Z\to \tau^-\mu^+\ \mbox{or}\ \tau^+\mu^-) &\leq & 1.9\cdot 10^{-5}\, ,
\label{BexpZlilj}
\end{eqnarray}
at $95\%$ CL. In addition, the Lagrangian (\ref{Znunu}) modifies the
invisible width of the $Z$ boson through the non-universal and
flavour-dependent $Z\nu_i\nu_j$ tree-level couplings. It is then easy
to obtain the branching ratio for the total invisible $Z$-boson
width, which is assumed to be caused mainly by $Z\to \nu_i\nu_j$
\begin{equation}
B(Z\to \nu\bar{\nu})\ =\ \frac{\alpha_w}{24\cos^2\theta_w}\,
\frac{M_Z}{\Gamma_Z}\sum\limits_{\nu_i,\nu_j}\, \left|\tilde{C}_{\nu_i
\nu_j}\right|^2. \label{BZnunu}
\end{equation}
On the other hand, an experimental analysis on the $Z$ pole
gives~\cite{PDG94}
\begin{equation}
1 - \frac{B(Z\to \nu\bar{\nu})}{B_{SM}(Z\to \mbox{invisible} )}\
\leq \ 1.31\cdot 10^{-2}\, ,\label{BexpZnunu}
\end{equation}
where $B_{SM} (Z\to \mbox{invisible} )$ is the SM prediction for the
invisible width of the $Z$ boson.
In Section 6, we will analyze the phenomenological impact of
the new-physics decay channels mentioned above on restricting
our model.

\subsection{ Universality violation at the {\em Z} peak}
\indent

Interesting limits on $R$-parity breaking, nonuniversal, diagonal $Zll$
couplings can be extracted from measurements of lepton universality
on the $Z$-boson pole. In order to impose constraints, we will adopt the
LEP observable based on leptonic $Z$-boson partial width
differences studied in~\cite{BKPS}
\begin{equation}
U_{br}^{(ll' )}\ =\ \frac{\Gamma(Z\to l^+l^-)
- \Gamma (Z\to l'^+l'^-)}{\Gamma(Z\to l^+l^-) + \Gamma (Z\to l'^+l'^-)}
\ =\ \frac{ |\tilde{A}^L_{ll}|^2\ -\ |\tilde{A}^L_{l'l'}|^2}{
|\tilde{A}^L_{ll}|^2\ +\ |\tilde{A}^L_{l'l'}|^2}\, ,\label{Ubr}
\end{equation}
where $l\ne l'$. A combined experimental analysis for
the observable $U_{br}$ gives~\cite{PDG94}
\begin{eqnarray}
|U_{br}^{(ll')}| \ \leq \ 5.0\cdot 10^{-3}\, ,\label{Ubrexp}
\end{eqnarray}
at 1$\sigma$ level, almost independent of the charged leptons $l$ and $l'$.
Another relevant observable involving leptonic asymmetries, which
has been analyzed in~\cite{BP}, is
\begin{equation}
\Delta{\cal A}_{ll'}\ =\ \frac{{\cal A}_l\ -\ {\cal A}_{l'}}{
{\cal A}_l\ +\ {\cal A}_{l'}}\ =\ \left(\frac{1}{{\cal A}_l^{(SM )}}\
-\ 1\right)\, U_{br}^{(ll')}\, ,\label{DAll'}
\end{equation}
where ${\cal A}_l^{(SM)}=0.14$ is the leptonic asymmetry predicted
theoretically in the SM.
In the last step of Eqs.~(\ref{Ubr}) and~(\ref{DAll'}), we have used the
fact that, to a good approximation, the tree-level coupling of the $Z$ boson
to right-handed charged leptons is universal in our minimal $R$-parity
violating SUSY model, {\em i.e.}\ $\tilde{A}^R_{ll}=\tilde{A}^R_{l'l'}$
as can be seen from Eq.~(\ref{AR}).
Considering the experimental upper bound on $U_{br}$ given in
Eq.~(\ref{Ubrexp}), Eq.~(\ref{DAll'}) furnishes the upper
limit
\begin{equation}
\Delta{\cal A}_{ll'} \ \le\  3.0\cdot 10^{-2}\, , \label{DAUbr}
\end{equation}
which is slightly below the present experimental sensitivity at
LEP~\cite{LEP} [$\Delta{\cal A}^{LEP}_\tau/{\cal A}^{(SM)}_l=0.07$, at
1$\sigma$] and Stanford Linear Collider (SLC)~\cite{SLD}
[$\Delta{\cal A}^{SLC}_e/{\cal A}_l^{(SM)}=0.04$, at 1$\sigma$].
It is also interesting to notice that the apparent difference of
$\Delta{\cal A}_{\tau e}\simeq - 10\%$ between the measured
leptonic asymmetries ${\cal A}^{SLC}_e$ and ${\cal A}^{LEP}_\tau$
cannot be predicted in our $\not\!\!\! R$ model, without invalidating
the inequality~(\ref{DAUbr}) at the same time.

\subsection{Decays {\boldmath $\mu\to e\nu\nu$} and
{\boldmath $\tau\to e\nu\nu$}}
\indent

Useful constraints can be obtained from possible deviations
of charged-current universality in $\tau$-lepton decays.
In fact, measures of such deviations can be defined
and straightforwardly be calculated as follows:
\begin{eqnarray}
R_{\tau e} &=& \frac{\Gamma (\tau\to e \nu\bar{\nu})}{
\Gamma (\mu\to e \nu\bar{\nu})}\ =\ R_{\tau e}^{SM}\,
\frac{\sum\limits_{\nu_i} \Big[ |\tilde{B}^L_{\tau\nu_i}|^2 +
|\tilde{B}^R_{\tau \nu_i}|^2 \Big]}{\sum\limits_{\nu_j}
\Big[ |\tilde{B}^L_{\mu\nu_j}|^2 + |\tilde{B}^R_{\mu \nu_j}|^2 \Big]}\, ,
\label{Rtaue}\\
R_{\tau \mu} &=& \frac{\Gamma (\tau\to \mu \nu\bar{\nu})}{
\Gamma (\mu\to e \nu\bar{\nu})}\ =\ R_{\tau\mu}^{SM}\,
\frac{\sum\limits_{\nu_i} \Big[ |\tilde{B}^L_{\tau\nu_i}|^2 +
|\tilde{B}^R_{\tau \nu_i}|^2 \Big]}{\sum\limits_{\nu_j}
\Big[ |\tilde{B}^L_{e\nu_j}|^2 + |\tilde{B}^R_{e \nu_j}|^2 \Big]}\,
.\label{Rtaumu}
\end{eqnarray}
In Eqs.~(\ref{Rtaue}) and~(\ref{Rtaumu}), the SM contributions
to the observables, $R_{\tau e}^{SM}$ and $R_{\tau\mu}^{SM}$,
have been factored out. Of course, deviations from the SM values
can also be induced by the $\lambda$-dependent interactions in
Eq.~(\ref{Llambda}). These observables are used to constrain
the couplings $\lambda_{ijk}$ as a function of the mass of the scalar
right-handed leptons~\cite{BGH}. To avoid excessive complication,
we assume that all $\lambda_{ijk}=0$ and focus our study mainly on
the phenomenological consequences originating from the $\varepsilon_i$
terms in the superpotential. Furthermore, experimental limits
related to the ratios $R_{\tau e}$ and $R_{\tau\mu}$ may be presented
in the following way~\cite{DAB}:
\begin{eqnarray}
1\ -\ \frac{R_{\tau e}}{R^{SM}_{\tau e}} &=& 0.040 \pm 0.024\,
,\label{Rexptaue}\\
1\ -\ \frac{R_{\tau\mu}}{R^{SM}_{\tau\mu}} &=& 0.032 \pm 0.024\,
,\label{Rexptaumu}
\end{eqnarray}
at 1$\sigma$ level. Constraints obtained from Eqs.~(\ref{Rexptaue})
and~(\ref{Rexptaumu}) on the parameters of our $\not\!\! R$ model
will be discussed in Section 6.

\subsection{Charged-current universality in pion decays}
\indent

Complementary to the physical quantities $R_{\tau e}$ and $R_{\tau\mu}$
are the constraints derived
from the ratio $R_\pi=\Gamma (\pi\to e\nu )/\Gamma (\pi\to \mu\nu )$
in the $\pi^-$ decays. $R_\pi$ is an observable that measures
possible deviations from charged-current universality in the $e-\mu$ system.
It is not difficult to obtain
\begin{equation}
R_\pi\ \ =\ R_\pi^{SM}\,
\frac{\sum\limits_{\nu_i} \Big[ |\tilde{B}^L_{e\nu_i}|^2 +
|\tilde{B}^R_{e \nu_i}|^2 \Big]}{\sum\limits_{\nu_j}
\Big[ |\tilde{B}^L_{\mu\nu_j}|^2 + |\tilde{B}^R_{\mu \nu_j}|^2 \Big]}\,
.\label{Rpi}
\end{equation}
In addition, the 1$\sigma$ experimental bound related to $R_\pi$ is
given by~\cite{DAB}
\begin{equation}
\frac{R_\pi}{R_\pi^{SM}}\ -\ 1 \ =\ 0.003\pm 0.003\, .\label{Rpiexp}
\end{equation}
It is again worth mentioning that similar deviations of $e-\mu$
universality can arise from the presence of $\lambda'$-dependent
couplings through the interaction Lagrangian~(\ref{Llambda'}).
In our analysis, we will assume that all $\lambda'_{ijk}=0$.
This may also be reflected by the fact that the current experimental
lower bound on the half-lifetime of the $^{76}$Ge $0\nu\beta\beta$ decay
leads to the tight constraint~\cite{HKK}
\begin{equation}
\lambda'_{111}\  \le\ 3.9\cdot 10^{-4}\, \left( \frac{m_{\tilde{q}}}{100\
\mbox{GeV}}\right)^2 \left( \frac{m_{\tilde{g}}}{100\ \mbox{GeV}}
\right)^{1/2},
\end{equation}
where $\tilde{q}$ ($\tilde{g}$) is the scalar quark (gluino).

\subsection{KARMEN anomaly}
\indent

Recently, the KARMEN collaboration, which operates at RAL, has
reported an anomaly~\cite{KARMEN} in the time-dependence of
decay spectra coming from stopped pions. To account for the
KARMEN anomaly, one can make the plausible assumption that a new
massive weakly-interacting particle, say $x$, is produced in the
pion decays, {\em i.e.}\ $\pi^+\to \mu^+ x$~\cite{KARMEN,BPS}. The
mass of this hypothetical particle should be $m_x\simeq 33.9$ MeV,
since it should explain the apparent $\sim 2\sigma$ bump present in
the time distribution of decaying muon events, which should normally
fall off exponentially.
This experimental peak occurs with a time delay of 3.6~$\mu$sec
after all pulsed pions have promptly decayed.

A recent study~\cite{BPS} suggests that the $x$ particle
should have similar features with those of a neutrino, but it cannot be
the $\nu_\tau$, because $m_{\nu_\tau}< 31$ MeV at 95$\%$ CL~\cite{PDG94},
or another predominantly-isodoublet neutrino, without affecting limits coming
from the supernova 1987A. The authors
in~\cite{BPS} further advocate that a mainly-sterile neutrino scenario
could, in principle, be compatible with all constraints ---both terrestrial and
astrophysical---, since the production of $x$ particles both in supernova
and in the early universe could then be suppressed. Although
in our $\not\!\! R$ model the coupling mixing matrices describing the
charged- and neutral-current interactions differ crucially from
usual singlet-neutrino scenarios~\cite{ZPC}, the above discussion
is still valid and translates into the requirement that one neutralino
state, {\em e.g.}\
$\chi$, should be light, having a mass $m_\chi=m_x$. Assuming
that the KARMEN anomaly gets resolved by the decay
$\chi\to e^-e^+\nu$, we have for the Majorana fermion $\chi$~\cite{BPS}
\begin{eqnarray}
|\tilde{B}^L_{e\chi}|\, |\tilde{B}^L_{\mu\chi}| &\simeq & 0.6\cdot 10^{-6}\, ,
\nonumber\\
|\tilde{B}^L_{e\chi}| &\stackrel{\displaystyle <}{\sim} & 2.5\cdot 10^{-4}\, ,
\nonumber\\
|\tilde{B}^L_{\mu\chi}| &\stackrel{\displaystyle <}{\sim} & 4.5\cdot 10^{-2}\,
{}.
\label{KARMENmix}
\end{eqnarray}
The bounds presented in Eq.~(\ref{KARMENmix}) are obtained from a number
of phenomenological requirements, such as the absence of a correction
to the Michael $\rho$ parameter in $\mu\to e\nu\nu$, negligible
decay events in neutrino beams, no anomalous contributions to
$\pi\to e\chi$, limits from neutrinoless double-$\beta$
decays, {\em etc}.

Because of the large number of parameters existing in our model,
it appears not difficult to accommodate the upper limits and
relations given in Eq.~(\ref{KARMENmix}). However, the soft-SUSY
breaking parameters in our model have to satisfy the following
hierarchy scheme:
\begin{eqnarray}
M\, (=2M') &\stackrel{\displaystyle >}{\sim}& 500\ \mbox{GeV}\, ,\nonumber\\
\mu &\stackrel{\displaystyle <}{\sim}& 30\ \mbox{MeV}\, ,\nonumber\\
\vec{\varepsilon}  & \sim & \frac{\mu}{v_1}\vec{w}\, ,\label{KARMENsusy}
\end{eqnarray}
which is mainly prescribed by the fact that $m_\chi=33.9$~MeV.
{}From Eq.~(\ref{KARMENsusy}), we find that only SUSY models with
a $\mu$ at the scale of 10 MeV have a chance to account
for the KARMEN anomaly.
Similar $R$-parity broken SUSY models were also discussed in
Ref.~\cite{EGJRV}. Adapting the results of~\cite{EGJRV},
one can estimate that for $\tan\beta=1$ and $w_\tau<60$~GeV,
$B(Z\to \chi \chi)\simeq w_\tau^4/(3v^4)<1.\ 10^{-3}$ in compliance
with the LEP bound on invisible $Z$-boson decays in
Eq.~(\ref{BexpZnunu}).
However, such light-$\mu$ scenarios may encounter the
known $\mu$ hierarchy problem, where $\mu\sim M_{Pl}$
as derived naively from supergravity. Even though one
could invoke the Guidice--Masiero mechanism~\cite{GM}
to obtain a value of $\mu$ at the electroweak scale,
the small value of $\mu={\cal O}(10)$ MeV would, however,
require an additional unnatural suppression of the
gravitational couplings in the K\"ahler potential.

\subsection{Cosmological and astrophysical constraints}
\indent

The minimal SUSY model with explicit $R$ nonconservation contains
lepton-number violating interactions that can wash out any primordial
BAU generated at the GUT scale via the $B+L$-violating sphaleron
interactions~\cite{GTH,NM,KRS},
which are in thermal equilibrium above the critical temperature of the
electroweak phase transition~\cite{ADD}. Sphalerons generally conserve
the individual quantum numbers $B/3-L_i$~\cite{CDEO,DR}. In particular, it
has been shown in~\cite{DR} that
if only one separate lepton number is preserved in thermal equilibrium
({\em e.g.}\ $L_i$) and finite masses for the charged
leptons are taken into account in the analysis of chemical potentials, this is
then sufficient to protect any primordial excess in $L_i$, which can be
converted later on, via sphalerons, into the observed BAU. For our purposes,
we will assume that only one separate lepton number is conserved each time
in the full Lagrangian, when low-energy experiments are
considered. For definiteness, in our numerical analysis we will consider
that either $w_e=\varepsilon_e=0$ or $w_\tau=\varepsilon_\tau=0$. Of
course, one can use a complementary restriction and put
$w_\mu=\varepsilon_\mu=0$,
which, however, will not alter our phenomenological constraints discussed
in Sections~5.1--5.5 in an essential way.

There is a great number of bounds coming from astrophysics,
such as those obtained from the dynamics of red giants and white dwarfs,
or the absence of a distorted spectrum of the $2.73^\circ$ K
blackbody radiation background~\cite{GGR}. However, we find
more worrying the severe limits derived from possible reheating effects
of a decaying massive neutral relic with $m_\nu\simeq 10-40$
MeV and especially those obtained from the primordial
nucleosynthesis~\cite{EGLNS,EMR}.
In particular, $\tau$-neutrino decays with a lifetime bigger than about
1~sec or so may increase the elemental ${}^4$He abundance by making it
incompatible with astrophysical observations.  Imposing the
latter constrain, we find
\begin{equation}
|\tilde{B}^L_{e\nu_\tau}|^2\ \stackrel{\displaystyle >}{\sim}\
10^{-4}\, \left(\frac{30\, \mbox{MeV} }{m_{\nu_\tau}}\right)^5\, ,
\label{Bastr}
\end{equation}
which is only applicable for $m_{\nu_\tau}
\stackrel{\displaystyle >}{\sim} 10-50$~MeV~\cite{EGLNS}.
In fact, the bound of Eq.~(\ref{Bastr}) is not so restrictive,
since it simply constrains only the mixing-matrix element
$V^l_{e\nu_\tau}>10^{-2}$ in Eq.~(\ref{BL}),
which is not excluded from solar neutrino oscillation scenarios.
In our analysis of laboratory observables, we sum up over all invisible
light neutrinos, so the unitary matrix $V^l$ becomes practically
redundant. Moreover, the nonobservation of a $\gamma$ ray burst from
the Solar Maximum satellite after the supernova 1987A neutrinos
were detected may point towards the fact that the $\tau$ neutrino mainly
decays inside the supernova core. This leads again to $\nu_\tau$ lifetimes
compatible with the approximate inequality of Eq.~(\ref{Bastr}).
Even though the predicted supernova luminosity will increase
in such a case, an allowed window of scenarios that maximally violate $L_\mu$
and $L_\tau$ may be present in the $\sim 3$ MeV neutrino-sphere~\cite{BBHH}.

As has also been pointed out by the authors in Ref.~\cite{BBHH},
there may exist viable cosmological models in which $\nu_\tau$ is
stable with a mass of ${\cal O}(10)$~MeV. Such a solution requires
an alteration of the standard cosmological picture by, {\em e.g.},
reheating the universe even after inflation to only a few MeV and
invoking low-temperature baryogenesis as well~\cite{DH}. Then, the resulting
$\nu_\tau$ may not overclose the universe but it can even constitute
the cold dark matter.

For neutrinos with $m_\nu\stackrel{\displaystyle <}{\sim} 0.1$~MeV,
the cosmological bound regarding their lifetimes, $\tau_\nu$,
is different. In fact, $\tau_\nu$ should not be larger the age
of the universe, {\em i.e.}\ $\tau_\nu \stackrel{\displaystyle >}{\sim}
10^{23} \displaystyle{\big(\frac{m_\nu}{1\, \mbox{eV}}\big)}$~sec~\cite{EMR}.
Furthermore, it is worth mentioning that radiative decays of massive
neutrinos with $0.1\stackrel{\displaystyle <}{\sim} m_\nu
\stackrel{\displaystyle <}{\sim} 10$~keV have also found some applications
in cosmology and astrophysics~\cite{DWS}. In this context, most noticeable
is probably the Gunn-Peterson test~\cite{GP}, {\em i.e.}, the search of
primordial elements in the intergalactic medium. There seems to be a
deficiency of neutral hydrogen and helium in the intergalactic
medium~\cite{GP}.
A source of photo-ionization of these elements might be a radiatively decaying
neutrino. Here, we simply comment on the fact that the mass range of neutrino
required for such an explanation is different from what is suggested by
the solar neutrino puzzle and the atmospheric neutrino problem. To ionize
singly ionized helium, $m_{\nu}$ should be bigger than 109~eV, since the
ionization potential is 54.4~eV. A recent investigation of this issue may
be found in~\cite{SKS}.

\setcounter{equation}{0}
\section{Numerical results}
\indent

In this section, we will present numerical predictions as well as
constraints on the basic parameters of our $\not\!\! R$ model,
which have been discussed in Section 5. Although there is a large number
of parameters that could vary independently, it is important to remark that
there exists a strong correlation between new-physics observables and light
neutrino masses. This seems to be a generic feature of most of the $R$-parity
broken SUSY models considered in the literature~\cite{BBHH,NRV}.
However, a novel feature of our minimal $\not\!\! R$ scenario is that
the size of the scalar-neutrino vev's and the $\varepsilon_i$ terms can,
in principle, be unconstrained. In fact, if $\vec{\Lambda}\simeq 0$ in
Eq.~(\ref{31}), which is a form of alignment in the flavour space between
the vev's of the sneutrinos, $\vec{w}$, and the $\not\!\! R$ terms,
$\vec{\varepsilon}$, this condition alone is sufficient to evade upper limits
on the tau-neutrino mass for any value of the SUSY parameters $M$, $M'$,
$\mu$, and $\tan\beta$.

In order to understand how all new-physics interactions are proportional
to $\Lambda_i$ and hence depend on $m_{eff}$ [or $m_{\nu_\tau}$] in
Eq.~(\ref{m31}) [Eq.~(\ref{m35})],
we evaluate the mixing matrix $\xi$ defined in Eq.~(\ref{xi}). Thus, we have
\begin{eqnarray}
\xi_{i1} &=& \frac{g'M\mu}{2\, \mbox{det}M_4}\, \Lambda_i\, ,\nonumber\\
\xi_{i2} &=& -\, \frac{g cM\mu}{2\, \mbox{det}M_4}\, \Lambda_i\, ,\nonumber\\
\xi_{i3} &=& \frac{\varepsilon_i}{\mu}\, +\, \frac{(cg^2+g'^2)Mv_2}{4\,
\mbox{det}M_4}\, \Lambda_i\, , \nonumber\\
\xi_{i4} &=& -\, \frac{(cg^2+g'^2)Mv_1}{4\, \mbox{det}M_4}\, \Lambda_i
\, ,\label{xii}
\end{eqnarray}
for $i=1\, (e),\, 2\, (\mu)$, and $3\, (\tau)$. It is now easy to see that
only the elements $\xi_{i3}$ contain the dominant contributions characterized
by being {\em not} proportional to $\Lambda_i$. However, these contributions
vanish identically in the relevant expression
\begin{equation}
\delta_{\nu_i\nu_j}\,  -\, \tilde{C}_{\nu_i\nu_j}\ =\
[\tilde{\xi}^* (1-t_z) \tilde{\xi}^T ]_{\nu_i\nu_j},
\end{equation}
given in Eq.~(\ref{C}), since the element of the
diagonal matrix $(1-t_z)_{33}=0$. Consequently, in the limit
of vanishing $\tau$-neutrino mass, the invisible $Z$-boson width
predicted in our $\not\!\!\! R$ model will coincide with that found
in the SM.

Similar strong $m_{\nu_\tau}$ dependence occurs in the non-SM
part of the couplings $Zl_il_j$ and $Wl_i\nu_j$ via the mixing matrix
$\xi_{{}_L}$, which is given by
\begin{eqnarray}
(\xi^*_{{}_L})_{i1}=
\frac{g\, \Lambda_i}{\sqrt{2} (M\mu-\frac{1}{2}g^2v_1v_2)}\, ,
\nonumber\\
(\xi^*_{{}_L})_{i2}= \frac{\varepsilon_i}{\mu}\, -\,
\frac{g^2v_2\, \Lambda_i}{2\mu (M\mu-\frac{1}{2}g^2v_1v_2)}\, .\label{xiiL}
\end{eqnarray}
One can readily see that the dominant terms in $\xi_{{}_L}$
are contained in the
elements $(\xi_{{}_L})_{i2}$. However, in the $Zl_il_j$ coupling, the
new-physics
contributions are determined by
\begin{equation}
[\xi_{{}_L} (1-d) \xi^\dagger_{{}_L}]_{ij}\ =\ (\xi_{{}_L})_{i1}
(\xi^*_{{}_L})_{j1}\, ,
\end{equation}
and the elements $(\xi_{{}_L})_{i2}$ always get killed by the diagonal matrix
$(1-d)$. Thus, leptonic FCNC $Z$-boson decays and associated
universality-breaking effects are proportional to $\Lambda_i$ and are absent
if $\nu_\tau$ is massless. Moreover, we find that the non-SM contributions
present in the coupling $Wl\nu$ in Eq.~(\ref{BL}) are proportional to
\begin{equation}
\Big( -\xi_{{}_L}\xi_{{}_L}^\dagger\, -\,
\xi^*\xi^T\, +\, 2\xi_{{}_L}t_{{}_L}\xi^T\Big)_{ij}\ =\
-\, (\xi_{{}_L})_{i2}[(\xi^*_{{}_L})_{j2}-\xi_{j3}]\, -\,
\xi_{j3}[\xi^*_{i3}\, -\, (\xi_{{}_L})_{i2}]\, .\label{newW}
\end{equation}
Substituting Eqs.~(\ref{xii}) and (\ref{xiiL}) into Eq.~(\ref{newW}),
it is easy to verify that new-physics effects in
charged-current interactions are also very strongly correlated
with the light neutrino mass $m_\nu$.

For reasons mentioned above, we will work in the seesaw approximation
by keeping the mass of $\nu_\tau$ finite. For our illustrations, we will
consider the following modest $\not\!\! R$ SUSY scenarios:
\begin{equation}
\begin{array}{r|r|rrrr}
\mbox{Scenario}& (\mbox{type of line})
& \tan\beta & M\, \mbox{[GeV]} & \mu\, \mbox{[GeV]} &\varepsilon_\mu\
(\mbox{or}\ \varepsilon_\tau )\, \mbox{[GeV]} \\
\hline
\mbox{I}     &\mbox{(solid)}             & 1   & 50 & 500 & 0    \\
\mbox{II}    &\mbox{(dashed)}            & 1   & 50 & -50 & -0.5 \\
\mbox{III}   &\mbox{(dotted)}            & 1   & 100 & 200& 1    \\
\mbox{IV}    &\mbox{(dash--dotted)}      & 4   & 200 & 400& 2    \\
\end{array} \label{A-scenarios}
\end{equation}
where $M'=M/2$ and the type of line used in our plots is also
indicated.

First, we will study possible limits on the $\not\!\! R$
models in Eq.~(\ref{A-scenarios}) that may be derived by the
nonobservation of a muon decay into three electrons. In Fig.~1(a), numerical
predictions for $B(\mu^-\to e^-e^-e^+)$ as a function of $m_{\nu_\tau}$
are displayed for $w_\mu/w_e=1$.
The horizontal dotted line indicates the present experimental
limit. Fig.~1(a) also shows the strong quadratic dependence of
$B(\mu^-\to e^-e^-e^+)$ on $m_{\nu_\tau}$. In particular,
if $w_\mu$ and $w_e$ are comparable in size ({\em e.g.}, $w_\mu/w_e=1$),
this constraint is more severe. Qualitatively, we find
that
\begin{equation}
\frac{w_ew_\mu}{w_e^2+w_\mu^2}\, \frac{m_\nu}{M}
\ \stackrel{\displaystyle <}{\sim}\ 10^{-6}\, .\label{bound}
\end{equation}
Of course, this limit gets relaxed for large vev ratios $w_\mu/w_e$.
The bound derived from $B(\mu^-\to e^-e^-e^+)$ is more sensitive to
the soft-SUSY gaugino mass $M$. To be more precise, our analysis
yields the following upper limits on $m_\nu$:
\begin{equation}
\begin{array}{r|r}
\mbox{Scenario}& m_\nu\, \mbox{[MeV]} \\
\hline
\mbox{I}     & 0.20    \\
\mbox{II}    & 0.57    \\
\mbox{III}   & 0.43    \\
\mbox{IV}    & 0.89    \\
\end{array} \label{A-bound}
\end{equation}
at 90$\%$ CL. We also remark that $\tau$-lepton number
is assumed to be conserved so as to protect a primordial excess in $L_\tau$
from being erased by processes that are in thermal equilibrium.
It is then obvious that for scenarios with $w_\mu/w_e=1$ and $M=200$ GeV,
$m_{\nu_\tau}<0.9$ MeV. From~(\ref{A-bound}), we see that scenario I
gives a stronger limit than the experimental one on the mass of $\nu_\mu$,
which is currently $m_{\nu_\mu}<0.27$~MeV at 90$\%$ CL~\cite{PDG94}.
As the non-SM couplings depend crucially
on the $\tau$-neutrino mass, the less than 1 MeV upper bound on a massive
neutrino gives little chance to see new-physics effects in other observables.
However, if $\Delta L_e=0$ in the model, {\em i.e.}\ $w_e=\varepsilon_e=0$,
inequality~(\ref{bound}) is trivially fulfilled and the so-derived neutrino
mass bound does not apply any longer.

In Fig.~1(b), numerical estimates reveal that non-SM contributions
to the invisible $Z$-boson width are one order of magnitude smaller
than the present experimental sensitivity. As a result, experimental
searches for physics beyond the SM, based solely on neutrino counting at
the $Z$ peak, are bound to be inadequate to unravel the nature of our minimal
$\not\!\! R$ model.

{}From Fig.~2(a), it can be seen that our minimal $\not\!\! R$
model may predict universality-breaking effects via the observable
$U_{br}$ in excess of $10^{-3}$. Such new-physics phenomena might
be seen at LEP, if all the experimental data accumulated in the year
1995 are analyzed.

Furthermore, in Fig.~2(b), we give theoretical predictions for the
observable $R_\pi/R_\pi^{SM}-1$ given in Eq.~(\ref{Rpiexp}). Possible
deviations from lepton universality in charged-current interactions
turn out to be one order of magnitude smaller than those that
can be accessed in experiment. Also, beyond the realm of detection
are found to be possible violations of charged-current universality
in the decays $\tau\to e \nu\nu$ and $\mu\to e \nu\nu$, which are
measured by virtue of the physical quantities $R_{\tau e}$ and
$R_{\tau\mu}$. Theoretically, similar is predicted to be the
situation for the size of the FCNC $Z$-boson mediated decays, such as
$\tau^-\to \mu^- e^- e^+$ and $Z\to ll'$. More explicitly,
it is estimated that
\begin{eqnarray}
B(\tau^-\to \mu^- e^- e^+) &\stackrel{\displaystyle <}{\sim}&
1.\, 10^{-9}\, ,\nonumber\\
B(Z\to l^-l'^+\ \mbox{or}\ l^+l'^-) &\stackrel{\displaystyle <}{\sim}&
1.\, 10^{-8}\, ,\nonumber\\
1\, -\, \frac{R_{\tau e}}{R_{\tau e}^{SM}} &\stackrel{\displaystyle <}{\sim}&
1.\, 10^{-4}\, ,\nonumber\\
1\, -\, \frac{R_{\tau\mu}}{R_{\tau\mu}^{SM}} &\stackrel{\displaystyle <}{\sim}&
1.\, 10^{-4}\, .
\end{eqnarray}

There may also be other places where $R$-parity violation could
manifest its presence. Of course, if neutralinos are lighter than the
$Z$ boson, one could search for distinctive signatures caused by decays
of the form $Z\to \nu_\tau \chi^0$ or $\tau^\pm \chi^\mp$, where
$\chi^0$ and $\chi^\pm$ decay subsequently into two $b$-quark jets
accompanied by a large amount of missing mass~\cite{BBHH}. However,
if the production threshold of heavy neutralinos and charginos is above
the LEP centre of mass energy, one then has to rely on studies of
possible indirect non-SM signals via sensitive observables
devoid of ambiguities coming from the evaluation of hadronic matrix
elements, as those discussed in Sections 5.1--5.4.
In the same logic, $R$-parity violating effects may also be probed
in the $\nu_\mu e$ scattering, even though experimental data
do not impose very stringent constraints as compared to those resulting
from $B(\mu\to e^-e^-e^+)$~\cite{BGH,BGKLM}.
Since our minimal $\not\!\! R$ model
only modifies the leptonic sector, one may derive useful constraints
from atomic parity violation measurements of the effective `weak charge',
$Q_W$, of a heavy nucleus. In the case of $^{133}_{55}$Cs, one
has~\cite{BGKLM}
\begin{equation}
Q_W^{exp}(^{133}_{55}\mbox{Cs})-Q_W^{SM}(^{133}_{55}\mbox{Cs})\ =\
73.5\cdot [\tilde{\xi}_L (1-d) \tilde{\xi}_L^\dagger]_{11}\ \le\ 3.74
\, ,\label{QW}
\end{equation}
at 1$\sigma$. The above bound turns out to be rather weak when
compared to that derived from $B(\mu\to eee)$.
Finally, for reasons that have already been mentioned in Section 5.3,
possible limits obtained directly from forward-backward-asymmetry
observables similar to $\Delta{\cal A}_{ll'}$ are estimated to be much
weaker than those determined
by the universality-breaking parameter $U_{br}^{(ll')}$ in Eq.~(\ref{Ubr}),
and are therefore not taken into consideration here.

\section{Conclusions}
\indent

The minimal $R$-parity broken SUSY model contains bilinear
lepton-number-violating terms ($\varepsilon_i$), which cannot in general
be eliminated
by a re-definition of the superfields provided soft-SUSY breaking parameters
are simultaneously present in the superpotential. The consideration of these
$\varepsilon_i$ mass terms, which involve the chiral multiplets of the
left-handed leptons and the Higgs field with $Y=+1$, give rise naturally
to non-vanishing vev's, $w_i$, of the
scalar neutrinos after the spontaneous breaking of the gauge symmetry.
In particular, if the vectors $\vec{w}$ and $\vec{\varepsilon}$,
spanned in the flavour space, satisfy a kind of alignment relation,
$\vec{\Lambda}=0$, forced, {\em e.g.}, by some horizontal symmetry, the
afore-mentioned $w_i$ and $\varepsilon_i$ parameters are not restricted
by limits on the $\tau$-neutrino mass. Furthermore, constraints from primordial
nucleosynthesis and the observed BAU have been considered. Specifically,
to evade BAU constraints  has been sufficient to impose that at least
one separate leptonic number has to be conserved in our $\not\!\! R$ model,
{\em e.g.}\ $w_\tau=\varepsilon_\tau=0$ and $w_e=\varepsilon_e=0$.

Our main interest has been to investigate the phenomenological implications
of this novel $\not\!\!\! R$ model in the light of a number of terrestrial,
astrophysical, and cosmological constraints.
To be more concrete, we have considered a typical set
of $\not\!\! R$ models as is stated in (\ref{A-scenarios}) and
confronted it with results obtained from LEP, CLEO and other
experiments.  We have found that the resulting non-SM contributions to
the couplings $Z\nu\nu$, $Zll'$, and $Wl\nu$ show a strong correlation
with the $\tau$-neutrino mass and vanish in the massless limit. This
direct correlation between the size of $R$-parity-violating phenomena
and the magnitude of the neutrino mass appears to be a generic feature
of most of the $R$-parity broken models considered in the
literature~\cite{BBHH,NRV}. In our analysis, the most severe
constraint comes from $B(\mu\to eee)$ for $\not\!\! R$ scenarios,
where $\Delta L_\tau=0$, and $L_e$ and $L_\mu$ are maximally violated.
In this way, we have been able to set an upper bound on $m_{\nu_\tau}$
by means of Eq.~(\ref{bound}). For instance, for $M=\mu=2M_W$ and
$w_e=w_\mu$, we find that $m_{\nu_\tau}\stackrel{\displaystyle <}{\sim}
1$ MeV. Especially, for scenario I in Eq.~(\ref{A-scenarios}),
we have $m_\nu < 0.2$~MeV as has been given in Eq.~(\ref{A-bound}),
which is even tighter than the current experimental bound on the mass of
the $\mu$ neutrino.
The remaining observables leave the main bulk of the parameter
space unconstrained. The most encouraging prediction is obtained
for the universality-violating observable $U_{br}$, with
$U_{br}\stackrel{\displaystyle <}{\sim} 2.\, 10^{-3}$. Such phenomena
might be seen at LEP, when the analysis of all the data of the year 1995
is completed.

For our purposes, we need not study the combined effect of the trilinear
$R$-parity-violating couplings $\lambda$ and $\lambda'$, {\em i.e.}\
$\lambda_{ijk}=\lambda'_{ijk}=0$.
The reason is that the Yukawa couplings $\lambda_{ijk}$ and $\lambda'_{ijk}$
are not sufficient to explain possible new-physics phenomena that can be
shown up in certain low-energy processes and LEP observables,
such as $B(l^-\to l'^-l_1^-l^+_1)$, $B(Z\to ll')$, and $U_{br}$,
discussed in Sections~5.1--5.3. In this context, we
remark that the KARMEN anomaly can, in principle, be explained by
assuming the presence of a fourth light neutralino, even though an
unnaturally small value of $\mu={\cal O}(10)$~MeV may be required.

\vskip1cm
\noindent
{\bf Acknowledgements.} The authors gratefully acknowledge discussions
with Roger Phillips. M.N. would like to thank the theory group of
Rutherford Appleton Laboratory for hospitality extended to him during
a visit, when part of this work was done.
M.N. also gratefully acknowledges financial support by
the HCM program under EEC contract no. CHRY-CT 920026.

\newpage
\setcounter{section}{0}
\def\theequation{\Alph{section}.\arabic{equation}}
\begin{appendix}
\setcounter{equation}{0}
\section{Appendix}
\indent
For completeness, we present  that part of the
scalar potential (\ref{m9}) that contains the charged singlet fields
$E_i$. This also consists of a lepton-number conserving contribution
($V_+^{L}$) and a lepton-number violating one ($V_+^{\not L}$).
The former reads
\begin{eqnarray} \label{a1}
V_+^{L}&=&[\mu^2_{+ij}(E_i^* E_j) + \mbox{H.c.}] +
[\mu_{ij}(\phi_2^{\dagger}\varphi_i)E_j + \mbox{H.c.}]
+[\mu'_{ij}(\phi_1^{\dagger}\varphi_i)E_j + \mbox{H.c.} ]\nonumber \\
&&+\lambda(\sum_k E_k^* E_k)^2 +(\tilde{\kappa}_{jk} - \lambda \delta_{jk})
(\phi_1^{\dagger}\phi_1)(E_j^* E_k) +
\lambda(\phi_2^{\dagger}\phi_2)(E_k^{*} E_k) \nonumber \\
&&+ (\mu_{ijnm} - \lambda \delta_{in}\delta_{jm})(\varphi_i^{\dagger}
\varphi_n)(E_j^{*} E_m) + 4\tilde{\kappa}_{nmij}(\varphi_n^{\dagger}
\varphi_i)(E_m^{*} E_j)\, ,
\end{eqnarray}
where we have defined
\begin{eqnarray} \label{a2}
&& \lambda = {1 \over 2} g'^2\, , \nonumber \\
&& \mu_{ij} = \mu^* h_{ij}\, , \nonumber \\
&& \tilde{\kappa}_{jk}= h_{ij}^* h_{ik}=\tilde{\kappa}_{kj}^*\, ,
\nonumber \\
&& \mu_{ijnm}= h_{ij}^* h_{nm}= \mu_{nmij}^* \, ,\nonumber \\
&& \tilde{\kappa}_{nmij}=\lambda_{knm}^* \lambda_{kij}=
\tilde{\kappa}_{ijnm}^*\, .
\end{eqnarray}
In Eq.~(\ref{a2}), $h_{ij}$ and $\lambda_{ijk}$ are couplings from the
superpotential (\ref{m3}) and (\ref{m4}), respectively.
$\mu_{+ij}^2$ and $\mu'_{ij}$ are soft-SUSY breaking parameters
from Eq.~(\ref{m8}).

For the lepton-number-violating contribution, we obtain
\begin{equation} \label{a3}
V_+^{\not L}= -2i\tilde{\kappa}_{ijk}(\phi_1^T \tau_2 \varphi_j)(E_i^*
E_k) + \kappa'_{ijk}(\varphi_i^T \tau_2 \varphi_j)E_k + \mbox{H.c.}\, ,
\end{equation}
where
\begin{equation} \label{a4}
\tilde{\kappa}_{ijk} = h_{ni}^* \lambda_{njk}\, ,
\end{equation}
and $\kappa'_{ijk}$ is a soft-SUSY breaking parameter contained in
Eq.~(\ref{m8}).

By analogy with Eq.~(\ref{m20}), from $V_+^L + V_+^{\not L}$, we can derive
conditions for not having CP violation in this part of the potential.
These conditions are listed below
\begin{eqnarray} \label{a5}
&& \mu^*_{ijnm}\eta^*_{{}_{L_n}} \eta_{{}_{L_i}}  \eta^*_{{}_{+m}}
\eta_{{}_{+j}}= \mu_{ijnm}\, , \nonumber \\
&& \tilde{\kappa}^*_{jk} \eta^*_{{}_{+k}}  \eta_{{}_{+j}}=\tilde{\kappa}_{jk}
\, ,\nonumber \\
&&\mu^*_{ij}\eta_{{}_{2}}  \eta^*_{{}_{L_i}}  \eta^*_{{}_{+j}}=\mu_{ij}
\, ,\nonumber \\
&&\mu'^{*}_{ij}\eta_{{}_{1}}  \eta^*_{{}_{L_i}}  \eta^*_{{}_{+j}}=\mu'_{ij}
\, ,\nonumber \\
&&\tilde{\kappa}^*_{nmij}\eta^*_{{}_{L_n}} \eta_{{}_{L_i}}  \eta^*_{{}_{+m}}
\eta_{{}_{+j}}= \tilde{\kappa}_{nmij}\, , \nonumber \\
&&\tilde{\kappa}^*_{ijk}\eta^*_{{}_{1}} \eta^*_{{}_{L_j}}  \eta_{{}_{+i}}
\eta^*_{{}_{+k}}= \tilde{\kappa}_{ijk}\, , \nonumber \\
&& \kappa'^{*}_{ijk}\eta^*_{{}_{L_i}} \eta^*_{{}_{L_j}}  \eta^*_{{}_{+k}}
=-\kappa'_{ijk}\, ,
\end{eqnarray}
where summation convention is not implied.
In Eq.~(\ref{a5}), $\eta_{{}_{+j}}$ are the CP phases of the scalar fields
$E_j$, similar to the notation of Eq.~(\ref{m20}). In general, both sets,
(\ref{m20}) and (\ref{a5}), should not be viewed independently of one another.
For instance, using the equalities in Eqs.~(\ref{m20}) and (\ref{a5}),
one can derive
(no summation convention)
\begin{eqnarray} \label{a6}
&& \Im m(\lambda_6 \mu_{ij}\mu'^{*}_{nm}\tilde{\kappa}_{mj}\kappa_{in})=0
\, ,\nonumber \\
&& \Im m(\lambda_6 \mu^*_{ijnm} \mu_{nm}\mu'^{*}_{ij})=0
\, ,\nonumber \\
&& \Im m(\mu^*_{ijnm}\kappa_{ni}\tilde{\kappa}^*_{mj})=0\, ,
\end{eqnarray}
and many similar relations of this kind.

\end{appendix}

\newpage


\newpage

\centerline{\bf\Large Figure Captions }
\vspace{-0.2cm}
\newcounter{fig}
\begin{list}{\bf\rm Fig. \arabic{fig}: }{\usecounter{fig}
\labelwidth1.6cm \leftmargin2.5cm \labelsep0.4cm \itemsep0ex plus0.2ex }

\item $B(\mu^-\to e^-e^-e^+)$ and $1-
B(Z\to \nu\nu)/B_{SM}(Z\to\mbox{invisible})$
versus $\tau$-neutrino mass, for the scenarios given in
Eq.~(\ref{A-scenarios}). We indicate
the present phenomenological limit by an horizontal dotted line.

\item The observables $U_{br}$ and $R_\pi/R^{SM}_\pi -1$ as a function
of the $\tau$-neutrino mass, for $\not\!\! R$ models stated in
Eq.~(\ref{A-scenarios}).
Horizontal dotted lines denote the present experimental bounds.

\end{list}

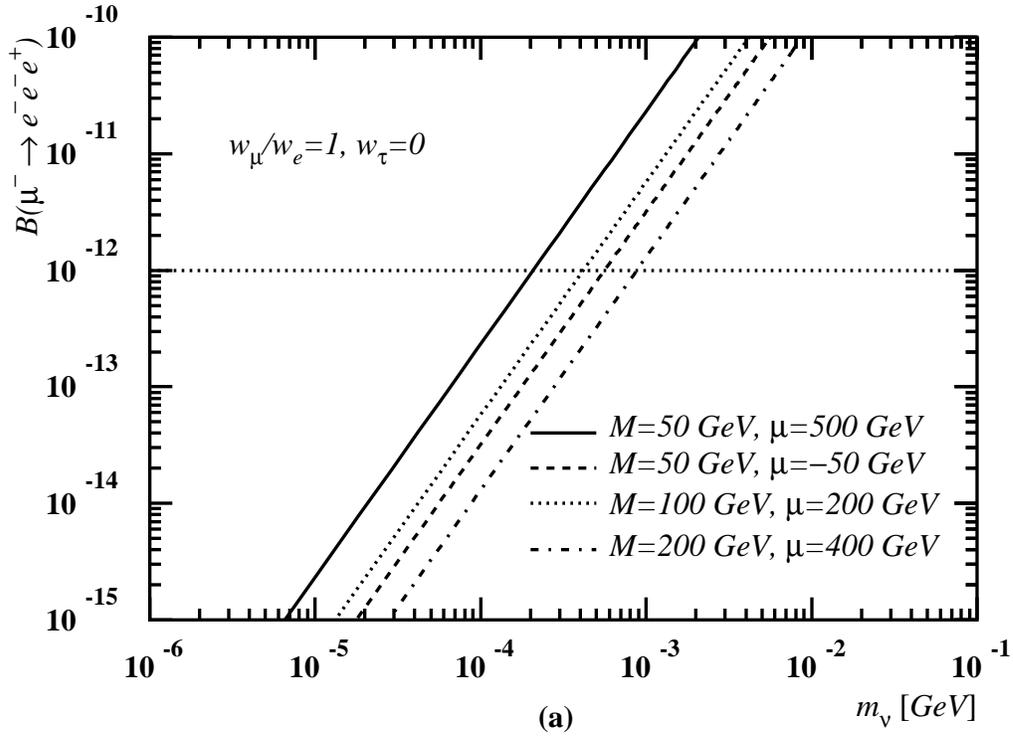
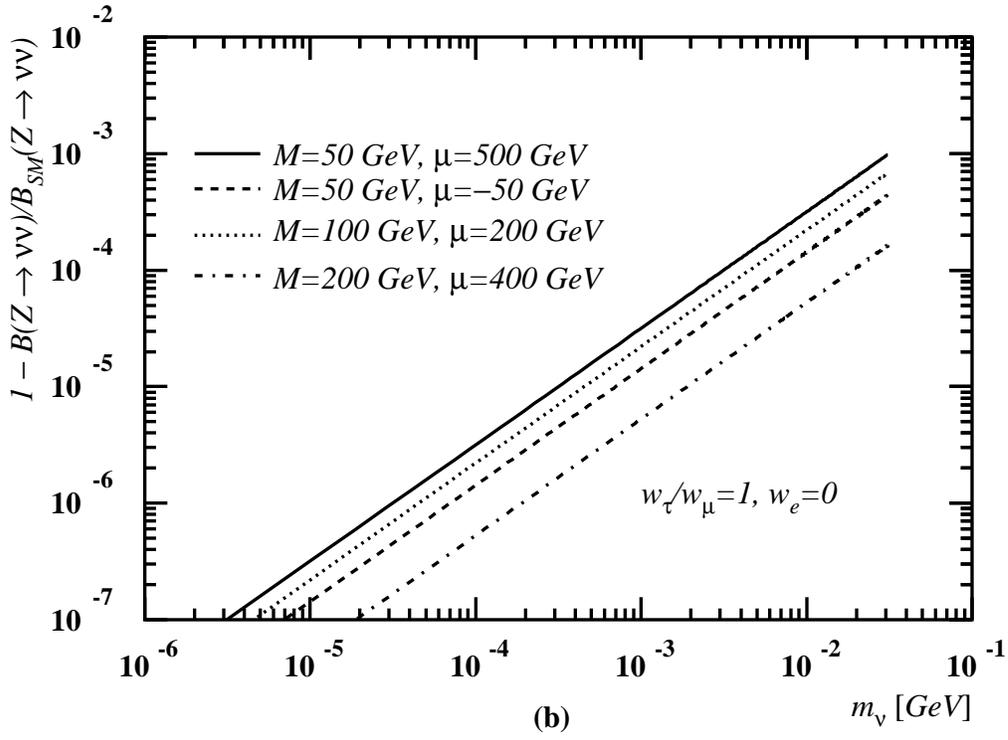
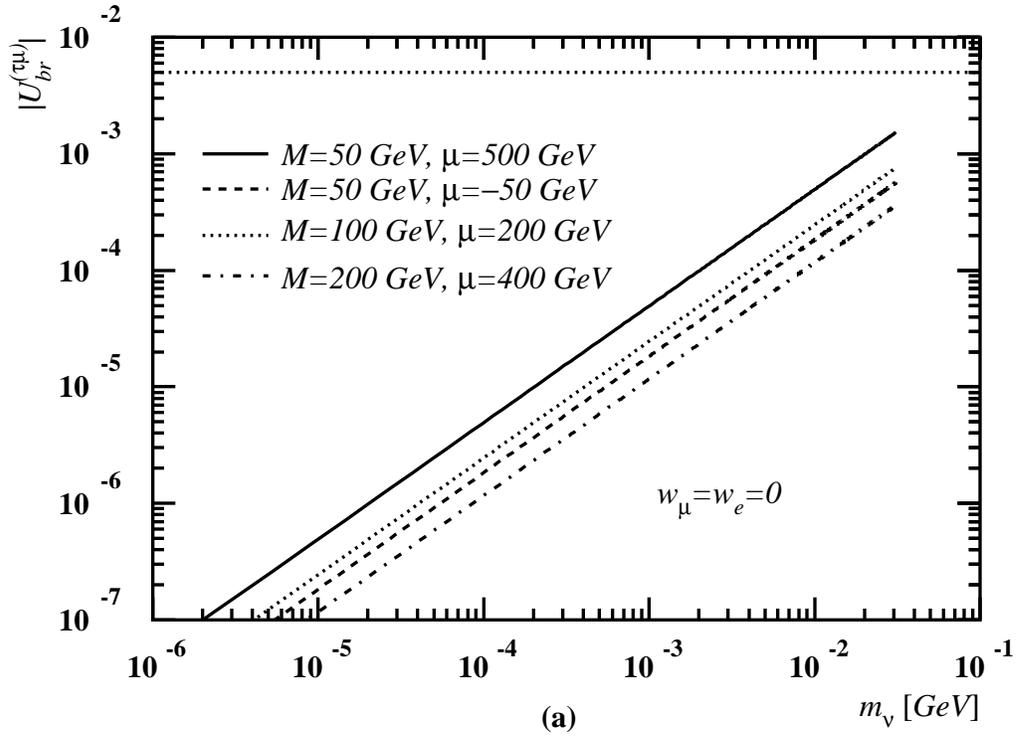
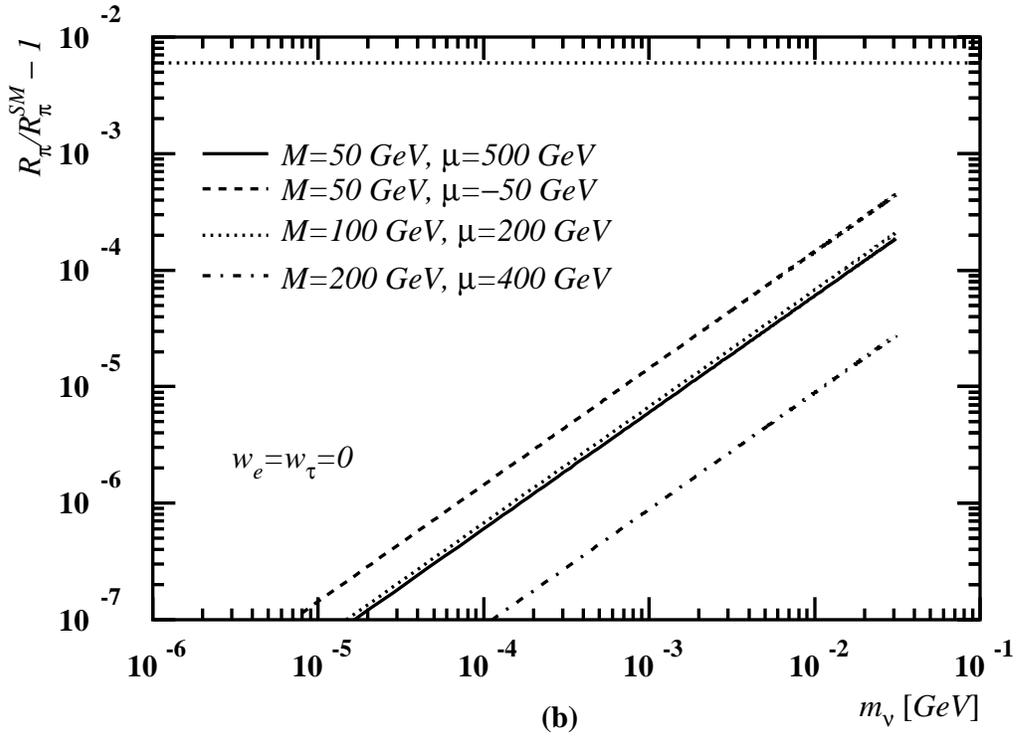
\end{document}